\documentclass[lettersize,journal]{IEEEtran}
\usepackage{amsmath,amsfonts}
\usepackage{bm}
\usepackage{algorithmic}
\usepackage{algorithm}
\usepackage{array}
\usepackage{textcomp}
\usepackage{tabularx}
\usepackage{stfloats}
\usepackage{url}
\usepackage{amssymb}
\usepackage{verbatim}
\usepackage{graphicx}
\usepackage{cite}
\usepackage{subfigure} 
\usepackage{epsfig}
\usepackage{epstopdf} 
\usepackage{bbding}

\usepackage{booktabs}
\usepackage{tabularx}
\usepackage[table]{xcolor}
\usepackage{multirow}

\usepackage{caption}
\newcolumntype{L}{>{\hspace*{-\tabcolsep}}l}
\newcolumntype{R}{r<{\hspace*{-\tabcolsep}}}
\definecolor{lightblue}{rgb}{0.93,0.95,1.0}

\newcommand{\figref}[1]{Fig. \ref{#1}}
\newcommand{\tabref}[1]{Table \ref{#1}}
\newcommand{\ra}[1]{\renewcommand{\arraystretch}{#1}}

\newcommand{\channelm}{\bm{h}_{m}}
\newcommand{\losm}{\bar{\bm{h}}_{m}}
\newcommand{\nlosm}{\tilde{\bm{h}}_{m}}

\UseRawInputEncoding

\begin{document}

\title{ CSI-tuples-based 3D Channel Fingerprints Construction Assisted by MultiModal Learning}

\author{
	Chenjie~Xie,~\IEEEmembership{Graduate Student~Member,~IEEE,}
	Li~You,~\IEEEmembership{Senior~Member,~IEEE,}\\
	Ruirong Chen, Gaoning He,
	Xiqi~Gao,~\IEEEmembership{Fellow,~IEEE}
	
	\thanks{			
	Part of this work was accepted by IEEE WCNC 2026 \cite{wcnc}. 
	
	Chenjie Xie, Li You, and Xiqi Gao are with the National Mobile Communications Research Laboratory, Southeast University, Nanjing 210096, China, and also with the Purple Mountain Laboratories, Nanjing 211111, China (e-mail: cjxie@seu.edu.cn, lyou@seu.edu.cn, xqgao@seu.edu.cn).
	
	Ruirong Chen and Gaoning He are with the Huawei Technologies Co., Ltd., Shenzhen 518129, China (e-mail: ruirongchen@huawei.com, hegaoning@huawei.com).			
}

}

%

\maketitle
\thispagestyle{empty}

\begin{abstract}
Low-altitude communications can promote the integration of aerial and terrestrial wireless resources, expand network coverage, and enhance transmission quality, thereby empowering the development of sixth-generation (6G) mobile communications.
As an enabler for low-altitude transmission, 3D channel fingerprints (3D-CF), also referred to as the 3D radio map or 3D channel knowledge map, are expected to enhance the understanding of communication environments and assist in the acquisition of channel state information (CSI), thereby avoiding repeated estimations and reducing computational complexity.
In this paper, we propose a modularized multimodal framework to construct 3D-CF.
Specifically, we first establish the 3D-CF model as a collection of CSI-tuples based on Rician fading channels, with each tuple comprising the low-altitude vehicle's (LAV) positions and its corresponding statistical CSI. In consideration of the heterogeneous structures of different prior data, we formulate the 3D-CF construction problem as a multimodal regression task, where the target channel information in the CSI-tuple can be estimated directly by its corresponding LAV positions, together with communication measurements and geographic environment maps. Then, a high-efficiency multimodal framework is proposed accordingly, which includes a correlation-based multimodal fusion (Corr-MMF) module, a multimodal representation (MMR) module, and a CSI regression (CSI-R) module.
Numerical results show that our proposed framework can efficiently construct 3D-CF and achieve at least 27.5\% higher accuracy than the state-of-the-art algorithms under different communication scenarios, demonstrating its competitive performance and excellent generalization ability. We also analyze the computational complexity and illustrate its superiority in terms of the inference time.
\end{abstract}


\section{Introduction}
As the sixth-generation (6G) mobile communications continue to evolve, massive demands for low-altitude applications have arisen across transportation, agriculture, and emergency services, catalyzing the vigorous development of low-altitude communications \cite{intro0, intro1, intro2, intro3, intro4}. Currently, low-altitude networks strive to provide real-time communication and navigation services for low-altitude vehicles (LAVs), facilitating the interoperability and coordination with terrestrial networks, delivering communication support for specific regions, and enhancing both network coverage and transmission quality \cite{intro0, intro4}. In the foreseeable future, low-altitude communications will further promote the synergistic integration of aerial and terrestrial resources, empowering a new paradigm for the development of mobile communications.

However, acquiring high-quality channel state information (CSI) in low-altitude communication systems remains an unresolved challenge, which significantly impacts the performance of low-altitude wireless transmission. Due to the spatially non-uniformly distributed scatterers \cite{intro5, in1}, the radio environments in low-altitude scenarios become more complicated, and the signal propagation characteristics vary significantly across different altitudes, rendering the acquisition of accurate CSI considerably challenging. On the other hand, LAVs in low-altitude airspace exhibit high mobility, constrained power consumption, limited payload, and restricted computing power \cite{intro5, intro6}, making the acquisition of real-time and efficient CSI more difficult.

Fortunately, channel fingerprints (CF), also referred to as the channel knowledge map (CKM) \cite{ckm} or the radio environment map (REM) \cite{rem}, have emerged as a novel approach to address the aforementioned challenge. By definition, CF is a site-specific database storing the user-location-related CSI. In terrestrial communication systems, 2D-CF has demonstrated its effectiveness in assisting the acquisition of channel information, which supports the resource management \cite{rm1}, beam selection \cite{bf1}, wireless positioning \cite{po1}, and other applications without repetitive estimations. Comparably, in low-altitude communication systems, 3D-CF will also directly provide the corresponding CSI based on the positions or trajectories of LAVs, thereby reducing pilot overhead, conserving wireless resources, and improving transmission efficiency. It is foreseeable that the introduction of 3D-CF will certainly bring a new perspective to the development of low-altitude communications.

Currently, several studies have commenced to explore approaches for high-efficiency 3D-CF construction.
For example, the authors in \cite{3D0} detected the number of radiation sources based on the path loss (PL) model and constructed a 3D spectrum map accordingly. Nevertheless, adopting such a one-size-fits-all PL model inevitably leads to substantial errors for 3D-CF construction due to its considerable variability across different altitudes.
To address the limitations of model-based approaches, data-driven methods have been extensively investigated.
For instance, the authors in \cite{kriging} leveraged LAV-based measurements to evaluate the interpolation algorithms, including nearest neighbor (NN), linear, inverse distance weighting (IDW), and ordinary Kriging, validating their feasibility for 3D-CF reconstruction. Gaussian process regression (GPR), sparse Bayesian learning, and compressed sensing (CS) were also developed to reduce the required number of samples in \cite{GPR, SBL, cs1, cs2, cs3}. Additionally, t-singular value decomposition (t-SVD), fiber sampling tensor decomposition (FSTD), block-term tensor decomposition (BTD), and other tensor-based algorithms were adopted to further enhance the computational efficiency by leveraging the smoothness prior of measurements in \cite{tensor2, tensor4, tensor1, tensor3}.
However, these pure data-driven methods are entirely environment-blind. In practice, wireless channels are profoundly influenced by the geographic environment through a complex interplay of signal reflection, diffraction, and scattering mechanisms \cite{envaware, envaware2}, especially in low-altitude communication systems where line-of-sight (LOS) paths are predominant.

To empower the geographic environment-assisted 3D-CF construction, researchers turn to machine learning (ML) for feasible solutions.
For instance, the authors in \cite{3ddnn} designed two deep neural networks (DNNs) to jointly reconstruct the 3D REM and its communication environment. Moreover, generative artificial intelligence (GenAI) was adopted for high-quality 3D-CF generation, supporting both radiation-aware and radiation-unaware scenarios with sparse spatial observations based on generative adversarial network (GAN) or diffusion model (DM) \cite{GAN2, GAN1, diff}.
However, these computer vision-based GenAI algorithms need to model 3D-CF as images, requiring a discretization for the target region where LAVs in the same grid (pixel) share the identical CSI (pixel values) \cite{cfcgn, jzz}. This assumption introduces some critical limitations. Firstly, the uniform discretization reduces the flexibility of 3D-CF. Unlike terrestrial networks, LAV trajectories in low-altitude airspace are highly communication-demand-driven \cite{lavtr}. Indiscriminate gridding merely results in a mismatch between 3D-CF resolution and communication demands while wasting computational resources. Secondly, the intra-grid CSI sharing induces significant errors for 3D-CF applications. Particularly when buildings exist within the grid, PL and channel shadowing for LAVs on opposite sides may differ substantially and cannot be represented by the same CSI. Thirdly, the grid-based 3D model triggers an exponential increase in data volume, resulting in high computational complexity and being unsuitable for practical deployment.

Multimodal learning (MML), a computer agent with intelligent capabilities such as understanding, reasoning, and learning, enables the integration of diverse data modalities for predictive tasks and offers a promising avenue to resolve the aforementioned issues of GenAI by transcending the image-only processing constraint. Currently, \cite{mmlbp1, mmlbp2, mmlbp3, mmlapp1, llm, mmlapp2, mmlapp3, mmlapp4, mmlapp4p, mmlapp5, llm2} have demonstrated enormous application potentials of MML-based CSI prediction in wireless communications. Particularly for 3D-CF construction in low-altitude systems, which involves more complex data modalities, including geographic environment maps, sparse communication measurements, and LAV coordinates, MML is expected to process and integrate the underlying information across these modalities, thereby offering novel solutions for 3D-CF construction and enhancing environmental awareness in low-altitude communication systems.

Motivated by the above discussions, we investigate the 3D-CF construction based on MML for low-altitude communication systems. Specifically, we encapsulate the LAV's coordinates and its corresponding channel information into a CSI-tuple, bypassing the discretization operation and directly fitting the mapping relationship between tuple elements to minimize the 3D-CF errors. Notably, the CSI stored in 3D-CF can be flexibly defined according to practical transmission requirements, such as the received signal strength (RSS), coverage, LOS probability, or even the channel covariance matrix and power angle spectrum (PAS). Then, we sufficiently consider the prior knowledge to assist the construction of 3D-CF, including geographic environment maps, sparse communication measurements, and LAV positions. Due to their heterogeneous data structures that can not be processed by a single-type network, we regard them as multimodal data and transform the 3D-CF construction problem into a multimodal regression task, thereby developing a modularized 3D-CF multimodal framework.
The main contributions of this paper can be summarized as follows:
\begin{itemize}
	\item Based on the ground-to-LAV Rician fading channel model, we propose a 3D-CF model that is more suitable for low-altitude communication systems. Specifically, the 3D-CF is conceptualized as a collection of CSI-tuples with each tuple comprising the LAV position and its corresponding channel information. This model enables the adaptive adjustments of CSI-tuples according to practical communication demands, thus making the flexible 3D-CF construction possible.
	\item Given the heterogeneous structures of different prior data, we formulate the 3D-CF construction problem as a multimodal regression task, where the target channel information in the CSI-tuple can be estimated directly by its corresponding LAV location, geographic environment maps, and measurement data.
	\item Based on the structural characteristics and internal relations of prior data, we propose a highly efficient modularized multimodal framework for 3D-CF construction. During the data processing stage, a correlation-based multimodal fusion (Corr-MMF) module and a multimodal representation (MMR) module are designed based on the relativity between communication environments and measurement data, which extract and learn features of the CSI distribution in horizontal and vertical directions, respectively. In the CSI estimation phase, we align these different features via embedding operations and design the channel state information regression (CSI-R) module to estimate CSI by leveraging LAV positions as conditional inputs, thereby recovering CSI-tuples and accomplishing the 3D-CF construction.
	\item We present numerical results to show that the proposed modularized multimodal framework achieves at least 27.5\% higher accuracy than state-of-the-art algorithms in 3D-CF construction. Experimental results also demonstrate its competitive performance in generalization capability and computational complexity. 
\end{itemize}

The rest of this paper is organized as follows. In Section \ref{S1}, we establish the ground-to-LAV channel model and 3D-CF model in low-altitude communication scenarios, and formulate the construction problem accordingly. Section \ref{framework section} elaborates on the structure of our proposed multimodal framework. Numerical results are presented in Section \ref{S3}. Finally, we conclude the paper in Section \ref{conclusion}.

\textit{Notations}: 
$\jmath = \sqrt{-1}$ denotes the imaginary unit.
$\bm{a}^{T} $ represents the transpose of vector $\bm{a}$ and $|| \textbf{A} ||_{F}$ is the Frobenius norm for matrix \textbf{A}. 
$\mathbb{C}^{M\times N\times K}$ denotes the $M\times N\times K$ dimensional complex tensor space and $\mathbb{R}^{3}$ represents the three-dimensional real space. 
$\mathbb{E}\{\cdot\}$ denotes the expectation operation.
$\mathcal{CN}(\bm{a}, \textbf{B})$ represents the complex Gaussian distribution with mean $\bm{a}$ and covariance $\textbf{B}$. 
The notation $\triangleq$ is used for definitions.

\section{System Model} \label{S1}
In this section, we introduce the channel model for ground-to-LAV links in low-altitude airspace. Then, a CSI-tuples-based 3D-CF model is established accordingly, which is highly flexible and does not rely on the spatial gridding. Based on the characteristics of 3D-CF, we formulate the construction problem as a multimodal regression task with the assistance of prior data.

\subsection{Channel Model} \label{channelmodel}
\begin{figure}[htbp]
	\centering
	\includegraphics[width=0.97\linewidth]{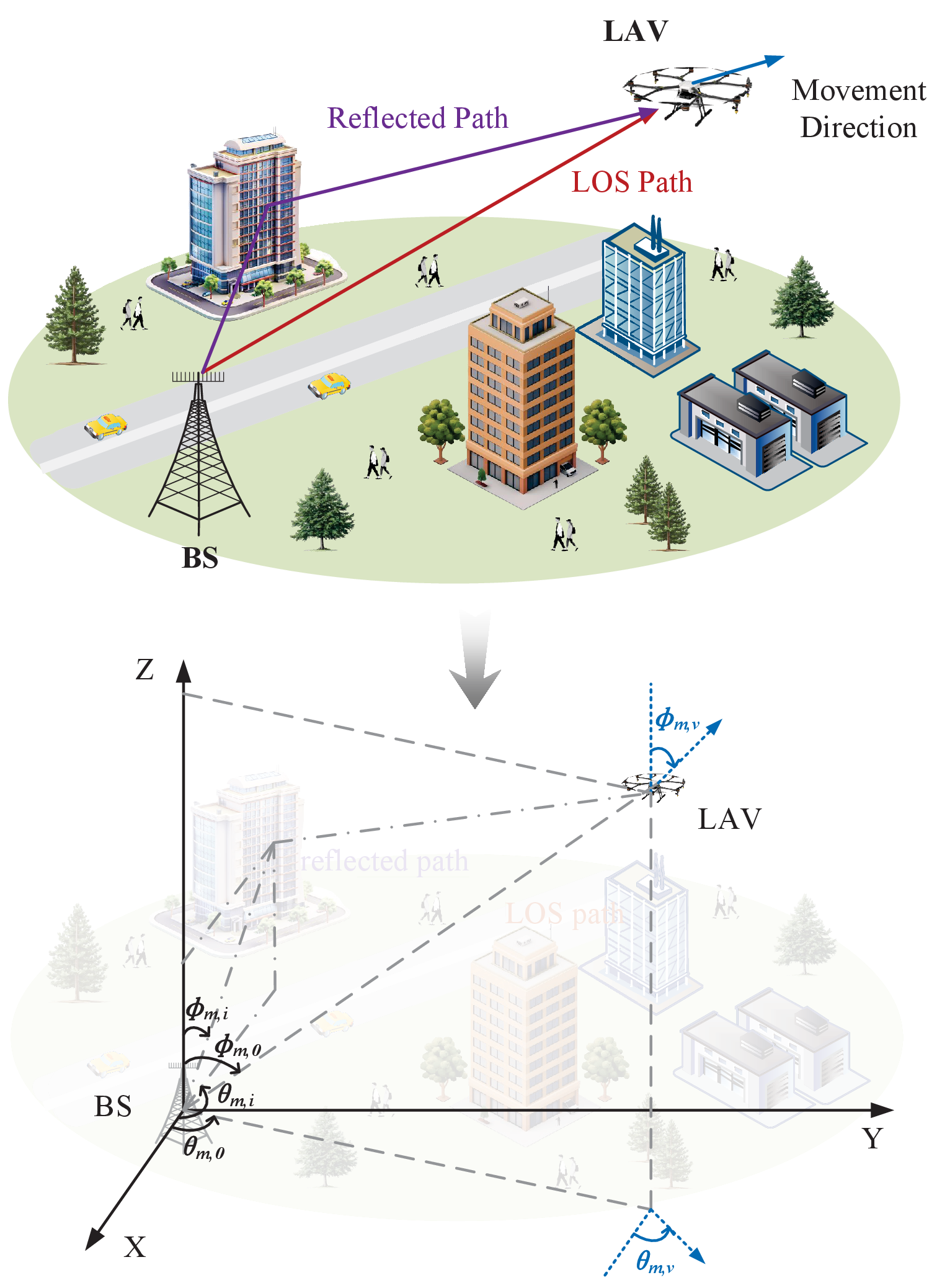}
	\caption{A typical ground-to-LAV communication scenario, where all possible channel components include one LOS path and several reflected paths. Its propagation geometry takes the BS as the origin.}
	\label{model}
\end{figure}

As illustrated in \figref{model}, we consider a ground-to-LAV system in low-altitude airspace, where the base station (BS), positioned at $(x, y, 0)$, is equipped with a uniform linear array (ULA) comprising $N_{\rm BS}$ antenna elements \cite{in2}. For simplicity, each LAV in the target region employs a single antenna and moves in a constant velocity in a time interval of interest.
To better characterize the ground-to-LAV links in target areas, we assume that each channel includes a line-of-sight (LOS) path and several reflected paths, both contributing to the received signal for a specific LAV \cite{model1}. By adopting the correlated Rician fading channel, the downlink (DL) channel between the BS and the $m$-th LAV over the $n$-th symbol can be modeled as \cite{model1, model2}
\begin{equation}
	\channelm[n] = \sqrt{\beta_{m}} \left( \losm[n] + \nlosm[n]\right),
	\label{cmo}
\end{equation}
where $\beta_{m}$ represents the large-scale channel fading coefficient, $\losm[n]$ and $\nlosm[n]$ denote the LOS component and NLOS component, respectively.

For the LOS component $\losm[n]$, define $K$ as the Rician factor and we have \cite{model2, core, modelad, modelad2}
\begin{equation}
	\losm[n] = \sqrt{\dfrac{K}{K+1}} \bm{\alpha}(\phi_{m, 0}, \theta_{m, 0}) e^{\jmath( 2 \pi \nu_{m} \bm{\xi}_{m, 0} T_{s} n + \varphi_{m, 0} )  },
	\label{los}
\end{equation}
where $\nu_{m}$ is the Doppler shift, $T_{s}$ is the system sampling duration, $\varphi_{m, 0}$ is the phase shift for LOS component, and $\bm{\xi}_{m, 0} \triangleq \bm{v}_{m} \bm{k}_{m, 0}^{T}$.
$\bm{v}_{m}$ is the unit velocity vector with an elevation angle $\phi_{m, v}$ and an azimuth angle $\theta_{m, v}$, which is given by
\begin{equation}
	\bm{v}_{m} = [ \cos(\theta_{m, v})\sin(\phi_{m, v}), \sin(\theta_{m, v})\sin(\phi_{m, v}), \cos(\phi_{m, v})].
	\label{v}
\end{equation}
$\bm{k}_{m, 0}$ is the unit wave vector with an elevation angle $\phi_{m, 0}$ and an azimuth angle $\theta_{m, 0}$, which is given by \cite{qmy}
\begin{equation}
	\bm{k}_{m, 0} = [ \cos(\theta_{m, 0})\sin(\phi_{m, 0}), \sin(\theta_{m, 0})\sin(\phi_{m, 0}), \cos(\phi_{m, 0})].
	\label{k}
\end{equation}
$\bm{\alpha}(\phi_{m, 0}, \theta_{m, 0})$ is the steering vector with an elevation angle $\phi_{m, 0}$ and an azimuth angle $\theta_{m, 0}$, which can be expressed as \cite{yyq}
\begin{equation}
	\bm{\alpha}(\phi_{m, 0}, \theta_{m, 0}) = [ 1, e^{\jmath 2\pi \frac{d_{m}}{\lambda_{c}} \zeta_{m, 0}}, \dots,  e^{\jmath 2\pi (N_{\rm BS}-1) \frac{d_{m}}{\lambda_{c}} \zeta_{m, 0}}],
\end{equation}
where $d_{m}$ is the inter-antenna spacing, $\lambda_{c}$ is the wavelength, and $\zeta_{m, 0} \triangleq \cos(\theta_{m, 0})\sin(\phi_{m, 0})$. 

For the NLOS component $\nlosm[n]$, define $L$ as the number of NLOS paths, we have \cite{core, ass1, ass2}
\begin{equation}
	\nlosm[n] = \sqrt{\dfrac{1}{K+1}} \sum_{l=1}^{L} \dfrac{\bm{\alpha}(\phi_{m, l}, \theta_{m, l})}{\sqrt{L}} e^{\jmath( 2 \pi \nu_{m} \bm{\xi}_{m, l} T_{s} n + \varphi_{m, l} )  },
	\label{nlos}
\end{equation}
where $\bm{\xi}_{m, l}\triangleq \bm{v}_{m} \bm{k}_{m, l}^{T}$ with $\bm{v}_{m}$ and $\bm{k}_{m, l}$ similar to (\ref{v}) and (\ref{k}), respectively. Assume that $\{\phi_{m, l}\}_{l=1}^{L}$, $\{\theta_{m, l}\}_{l=1}^{L}$, and $\{\varphi_{m, l}\}_{l=1}^{L}$ are independent random variables, then, according to the central limit theorem \cite{clt}, when $L$ tends to infinity, $\nlosm[n]$ will approximate a zero-mean complex Gaussian random process, i.e., $\nlosm[n]\sim\mathcal{CN}(0, \bm{\Lambda}_{m})$, where $\bm{\Lambda}_{m}$ represents the positive semi-definite spatial covariance matrix of the NLoS components for the $m$-th LAV \cite{model1, model2, core, ass1, ass2}. 

Based on the analysis of (\ref{los}) and (\ref{nlos}), the channel model between the BS and the $m$-th LAV over the $n$-th symbol can be expressed as $\channelm[n] \sim \mathcal{CN}(\textbf{H}, \textbf{R})$ \cite{model1, model2, core}, where mean $\textbf{H} = \sqrt{\beta_{m}}\losm[n]$ and covariance $\textbf{R}=\beta_{m}\bm{\Lambda}_{m}$. In accordance with this channel model, the RSS can be written by
\begin{equation}
	g_{m} = P_{\rm BS} ||\channelm[n] ||_{F}^{2},
	\label{gm}
\end{equation} 
where $P_{\rm BS}$ denotes the transmit power.

\subsection{3D-CF Model and Problem Formulation} \label{3dcfmodel}
Based on the channel model presented in Section \ref{channelmodel}, we next introduce the 3D-CF model for LAVs in low-altitude airspace and then formulate the construction problem.

\subsubsection{3D-CF Model}
Traditional 3D-CF model typically partitions the target area into grids, where all receivers within the same grid share the identical channel information, thereby converting CF into an image \cite{jzz, cfcgn}. 
In contrast, we model 3D-CF as a collection of CSI-tuples $\{(\mathcal{X}, \Omega)\}$, where $\mathcal{X}$ represents the LAV coordinates array and $\Omega$ denotes its associated channel information.
On one hand, for any LAV in low-altitude airspace, we can always locate its corresponding CSI-tuple in 3D-CF and obtain the accurate channel information, thereby reducing errors induced by spatial gridding. On the other hand, the collection of CSI-tuples can be dynamically adjusted or reconstructed according to terminal density, low-altitude traffic load, and spatial utilization, maximizing the 3D-CF flexibility while minimizing unnecessary computational overhead in practical applications.

In accordance with the channel model in Section \ref{channelmodel}, we define the RSS in (\ref{gm}) as the target channel information stored in 3D-CF, i.e., $\Omega=g_{m}$. Therefore, the collection of CSI-tuples is ultimately expressed as
\begin{equation}
\mathcal{G} = \{ (\mathcal{X}_{m}, \Psi(\mathcal{X}_{m})) | \Psi: \mathcal{X}_{m}\in \mathbb{R}^{3}\to g_{m} \},
	\label{3dcf}
\end{equation} 
which is our proposed 3D-CF model. Note that we define $\Omega$ as RSS solely for the convenience of elucidating the model, task, and methodology. In practice, $\Omega$ can be defined as different channel information according to practical communication requirements, such as PL, delay, Doppler shift, or even the channel covariance matrix, and the optimal beam indices, to match diverse applications.

\subsubsection{Problem Formulation}
Under the definition of (\ref{3dcf}), the problem of constructing 3D-CF is transformed into the task of exploring function $\Psi$, that is, finding a mapping relationship from the LAV location to its corresponding RSS. 

However, directly fitting $\Psi$ is extremely challenging due to the absence of distinct correlations between LAV location and its RSS. Consequently, we need to seek some prior information to facilitate the construction of $\Psi$.   
On one hand, the low-altitude geographic environment, which can be viewed as an image and conveniently captured by tools like RGB cameras, exerts a significant influence on the distribution of RSS. In terrestrial networks, geographic information has been proven to be effective and crucial in assisting the reconstruction of 2D-CF \cite{jzz, 2den1, 2den2, 2den3}. In low-altitude airspace, environmental effects on RSS are more pronounced due to the dominance of LOS paths in ground-to-LAV channels. Therefore, the low-altitude geographic environment, denoted as $\mathcal{E}$, is one of the essential prior information to facilitate the construction of $\Psi$.
On the other hand, measurable CF sampling data near the ground, denoted as tensor $\mathcal{G}_{\rm gro}$, can also serve as the prior information, as they partly reflect the signal propagation characteristics and reveal the underlying relationship between terminal locations and RSS.
Based on $\mathcal{E}$ and $\mathcal{G}_{\rm gro}$, the mapping relationship $\Psi$ can be rewritten as
\begin{equation}
	\Psi: (\mathcal{X}_{m}, \mathcal{E}, \mathcal{G}_{\rm gro}) \to g_{m}, \mathcal{X}_{m} \in \mathbb{R}^{3}.
	\label{mr}
\end{equation}

In practice, it is challenging to derive a feasible analytical solution by traditional interpolation methods, hence, we employ a deep neural network $\Psi'$ to fit $\Psi$ in (\ref{mr}).
In particular, the network $\Psi'$ involves three modal variables as inputs: the low-altitude communication environment $\mathcal{E}$, which encompasses both horizontal and vertical information of all buildings, vegetation, and other structures in the target low-altitude airspace; CF sampling data $\mathcal{G}_{\rm gro}$, which indicate the signal propagation characteristics; and the LAV position $\mathcal{X}_{m}$. Therefore, we formulate the problem of fitting $\Psi$ by the network $\Psi'$, i.e., the construction of 3D-CF, as a multimodal regression task, which is given by
\begin{align}
	\mathop{\arg \min}\limits_{\Theta}  \ & \mathbb{E} \left\lbrace || \Psi' \left[ (\mathcal{X}_{m}, \mathcal{E}, \mathcal{G}_{\rm gro}); \Theta \right] - g_{m} ||_{F}^{2} \right\rbrace \label{Problem} \\
	{\rm s.t.} \ & g_{m} = \Psi(\mathcal{X}_{m}, \mathcal{E}, \mathcal{G}_{\rm gro}) \tag{\ref{Problem}{a}}, \\
	\ &  \mathcal{X}_{m}\in \mathbb{R}^{3} \tag{\ref{Problem}{b}},
\end{align}
where $\Theta$ is the trainable parameters for the network $\Psi'$.

\section{MultiModal Framework for 3D-CF Construction} \label{framework section}
\begin{figure*}[t]
	\centering
	\includegraphics[width=1\linewidth]{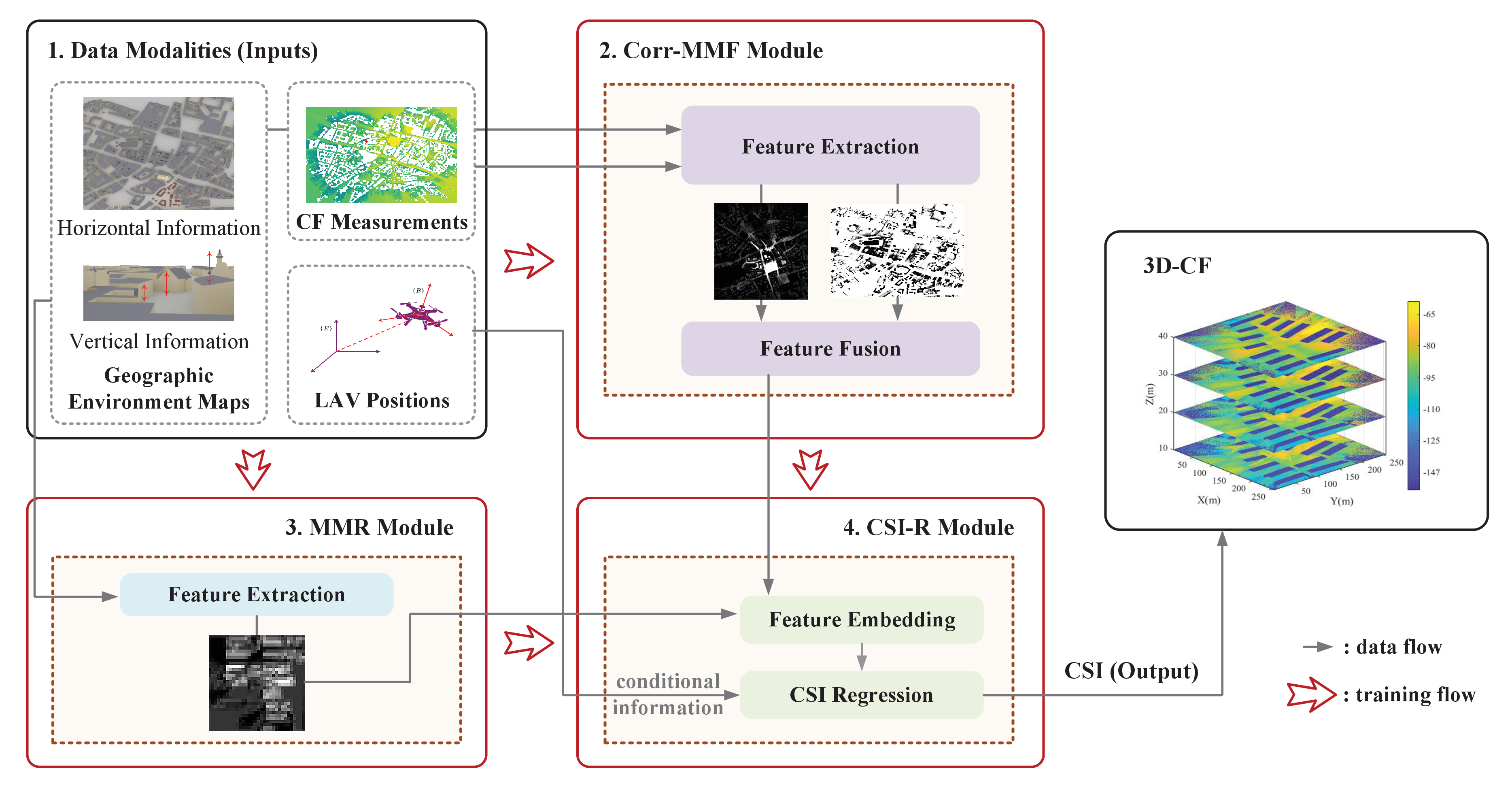}
	\caption{Diagram of the proposed 3D-CF MultiModal framework. This scheme includes three essential modules: the Corr-MMF module, the MMR module, and the CSI-R module. The first two modules are designed to extract features of CSI distribution and geographic environments in horizontal and vertical directions, respectively, and the third module is responsible for spacial CSI prediction and 3D-CF reconstruction.}
	\label{frame}
\end{figure*}

As analyzed in Section \ref{3dcfmodel}, mutual relationships among $\mathcal{X}_{m}$, $\mathcal{E}$, and $\mathcal{G}_{\rm gro}$ reveal the underlying patterns of spatial RSS distribution, hence, in this section, we develop a modularized multimodal framework to fit the mapping relationship $\Psi$ and construct 3D-CF.
As shown in \figref{frame}, our proposed 3D-CF Multimodal framework includes three essential modules: the Corr-MMF module, the MMR module, and the CSI-R module, where the first two modules are designed to extract features of CSI distribution and geographic environments in horizontal and vertical directions, respectively, and the third module is responsible for spacial CSI prediction and 3D-CF reconstruction based on these features. Next, we will introduce them respectively.

\subsection{The Correlation-based MultiModal Fusion (Corr-MMF) Module}
In conventional multimodal learning tasks, data from two or more media often exhibit strong correlations. Since their structural characteristics are heterogeneous, it is necessary to conduct a unified encoding and combination, known as MultiModal Fusion (MMF) \cite{mmf}. By definition, MMF is the process of extracting and integrating features from two or more media to perform the subsequent regression or classification \cite{mmf2}. It leverages the correlation and complementarity among different data to keep critical features and remove redundant ones, integrating various information into a stable multimodal representation \cite{mmf3}.

In our 3D-CF construction task, available CF measurements $\mathcal{G}_{\rm gro}$ near the ground exhibit a strong correlation with the horizontal geographic environment information $\mathcal{E}_{h}$ \cite{laerss, laerss2, laerss3}: on one hand, RSS in $\mathcal{G}_{\rm gro}$ reveal the possible distribution of buildings, vegetation, and other structures in the target area \cite{rsse}; on the other hand, $\mathcal{E}_{h}$ indicates the potential reflection, diffraction, scattering, and obstruction during signal propagation, thus affecting the distribution of RSS \cite{erss}. Consequently, exploring and fusing the correlated characteristics between $\mathcal{G}_{\rm gro}$ and $\mathcal{E}_{h}$ are crucial for the multimodal framework to learn 3D-CF patterns in the horizontal direction, which motivates our design of the Corr-MMF module. In particular, the low-dimensional feature representations extracted and fused by the Corr-MMF module must satisfy the following two criteria:

\textit{Criterion 1}: The low-dimensional feature representations should maximally preserve the critical information inherent to both $\mathcal{G}_{\rm gro}$ and $\mathcal{E}_{h}$;

\textit{Criterion 2}: The low-dimensional feature representations should effectively preserve the correlated information among $\mathcal{G}_{\rm gro}$ and $\mathcal{E}_{h}$.

\begin{figure}[htbp]
	\centering  
	\subfigure[Diagram of the proposed Corr-MMF module. The network includes a two-stages encoder, a latent space, and a virtual decoder. The terminal attention mechanism (TAM) and channel attention mechanism (CAM) are introduced in this two-stages encoder as well.]{
		\label{Corr-MMF module}
		\includegraphics[width=1\linewidth]{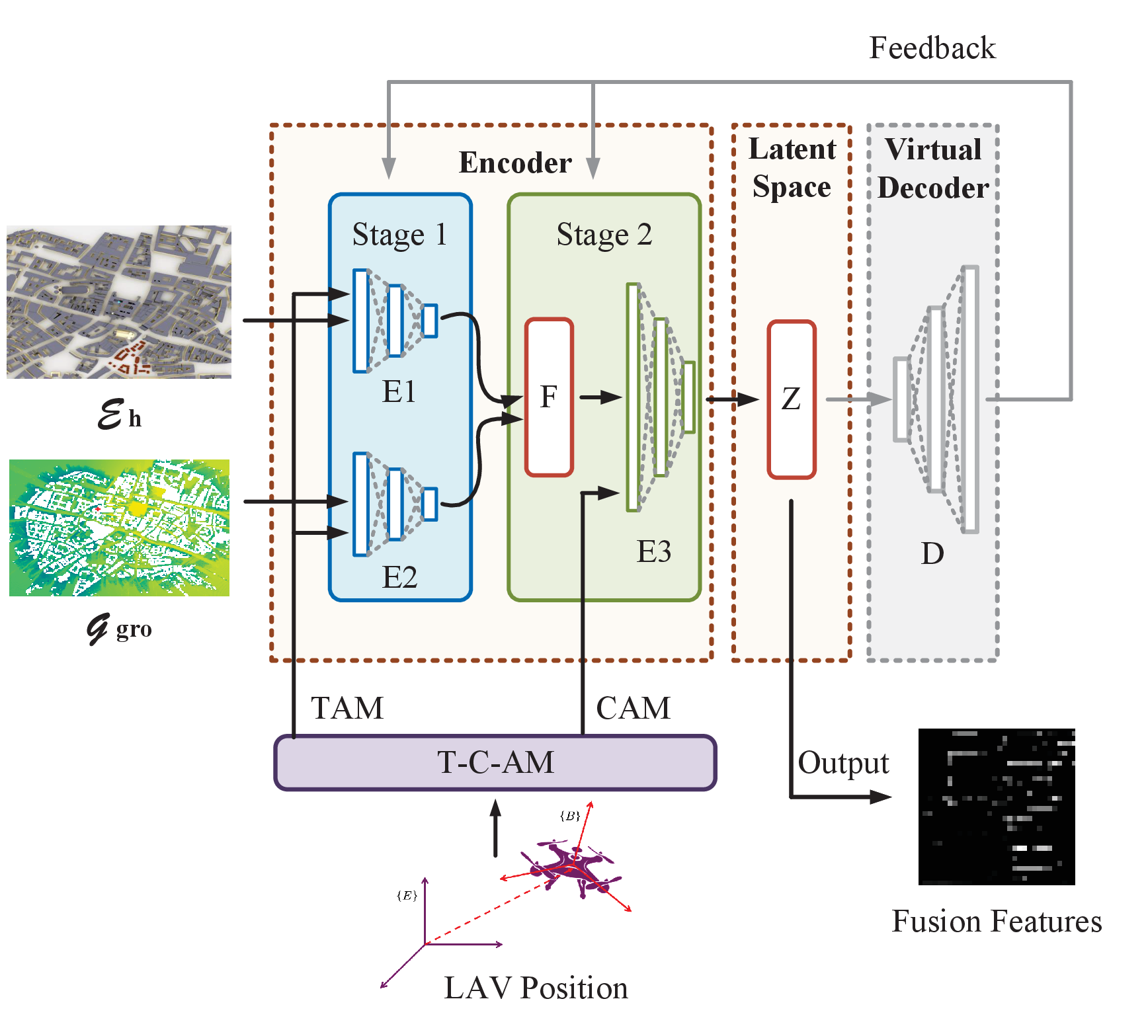}}
	
	\subfigure[Diagram of the channel attention mechanism (CAM).]{
		\label{CAM}
		\includegraphics[width=1\linewidth]{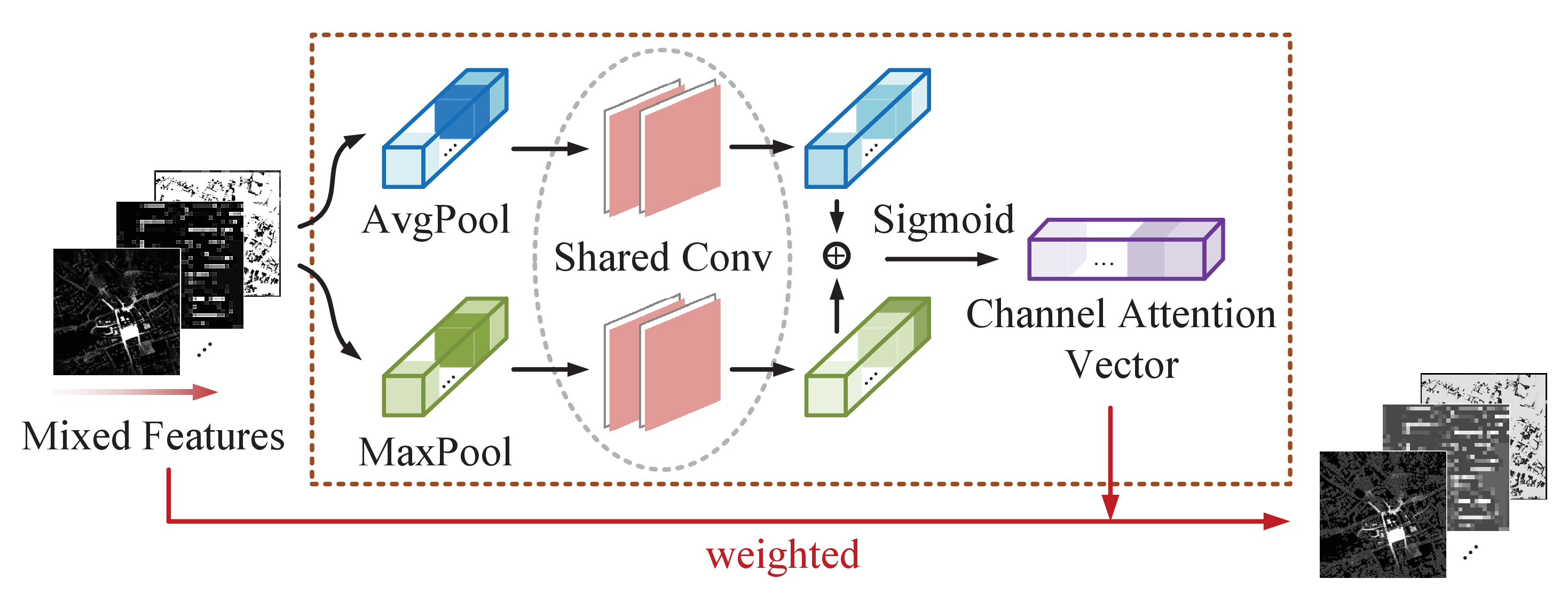}}                       
	\caption{Diagram of the Corr-MMF module and its CAM.} 
\end{figure}

\subsubsection{Network Design for Criterion 1}  \label{design_c1}
For Criterion 1, a workable structure is the feature extractor (encoder) used to implement the key information extraction for both $\mathcal{G}_{\rm gro}$ and $\mathcal{E}_{h}$. As shown in \figref{Corr-MMF module}, the encoder includes two stages: feature extraction and feature fusion.

During the feature extraction stage, two sub-encoders, $E_{1}$ and $E_{2}$, are employed to extract features of $\mathcal{G}_{\rm gro}$ and $\mathcal{E}_{h}$, respectively, eliminating data redundancy and achieving dimensionality reduction. Specifically, each of $E_{1}$ and $E_{2}$ consists of several convolutional layers, each of which is accompanied by a rectified linear unit (ReLU) to conduct the downsampling operation. To further speed up the convergences and simultaneously enhance the generalization performance of the network, we incorporate the batch normalization (BN) after each convolutional layer.

Furthermore, we observe that RSS of the specific LAV in 3D-CF exhibits a strong correlation with the channel information and geographical environment in its vicinity \cite{mask_envi}. Consequently, a terminal attention mechanism (TAM) is designed for $E_{1}$ and $E_{2}$ to focus on data blocks closer to the LAV. Specifically, we construct two Gaussian masks $M_{\mathcal{G}}$ and $M_{\mathcal{E}}$, with their dimensions identical to $\mathcal{G}_{\rm gro}$ and $\mathcal{E}_{h}$, respectively. For the ($m,n$)-th element in $M_{\mathcal{G}}$ and $M_{\mathcal{E}}$, we have
\begin{equation}
	M_{\mathcal{G}}(m,n) = M_{\mathcal{E}}(m,n)= e^{-\frac{d_{m,n}^{2}}{2\sigma^{2}}},
\end{equation}
where $d_{m,n}$ denotes the distance between this ($m,n$)-th element and LAV position, and $\sigma^{2}$ the variance of Gaussian distribution for $M_{\mathcal{G}}$ and $M_{\mathcal{E}}$. It is worth noting here that $M_{\mathcal{G}}$ and $M_{\mathcal{E}}$ in TAM will assign different weights to $\mathcal{G}_{\rm gro}$ and $\mathcal{E}_{h}$ at different spatial positions, thereby optimizing the process of feature extraction. Meanwhile, the Gaussian-distributed weights inherently maintain continuity and differentiability, facilitating the backward propagation computations for each sub-encoder.

Overall, outputs of the sub-encoders $E_{1}$ and $E_{2}$ can be written as
\begin{align}
	\mathcal{O}_{{\rm E}_{1}} &= {\rm BN}( {\rm ReLU}( {\rm Conv}( M_{\mathcal{G}} \odot \mathcal{G}_{\rm gro} ) ) ) \in \mathbb{C}^{h\times w\times c}, \\
	\mathcal{O}_{{\rm E}_{2}} &= {\rm BN}( {\rm ReLU}( {\rm Conv}( M_{\mathcal{E}} \odot \mathcal{E}_{h} ) ) ) \in \mathbb{C}^{h\times w\times c},
\end{align}
where $\odot$ represents the Hadamard product, $h, w, c$ are heights, widths, and channels of $\mathcal{O}_{{\rm E}_{1}}$ and $\mathcal{O}_{{\rm E}_{2}}$, respectively. 

During the feature fusion stage, we adopt an add layer to integrate $\mathcal{O}_{{\rm E}_{1}}$ and $\mathcal{O}_{{\rm E}_{2}}$. Then, the added feature $\mathcal{O}_{{\rm F}} = \mathcal{O}_{{\rm E}_{1}} + \mathcal{O}_{{\rm E}_{2}} \in \mathbb{C}^{h\times w\times c}$ is fed into a sub-encoder $E_{3}$ for further extraction of critical information, thereby completing the feature fusion. 
Note that the add layer here does not expand the number of channels in $\mathcal{O}_{{\rm F}}$, but rather exponentially enhances the feature informativeness they contain. Therefore, the subsequent sub-encoder $E_{3}$ must be capable of adaptively evaluating the importance of each channel to discern crucial features and emphasize them. 

To this end, we introduce a channel attention mechanism (CAM) to assign varying weights to different channels \cite{cbam}, as depicted in \figref{CAM}. Specifically, CAM employs an average pooling and a max pooling to separately obtain the global statistical information of each channel in mixed features $\mathcal{O}_{{\rm F}}$, denoted as ${\rm avgpool}{(\mathcal{O}_{{\rm F}})} \in \mathbb{C}^{1\times 1\times c}$ and ${\rm maxpool}{(\mathcal{O}_{{\rm F}})} \in \mathbb{C}^{1\times 1\times c}$, respectively. Subsequently, shared convolutional layers with a kernel size of $1$ are utilized to convert ${\rm avgpool}{(\mathcal{O}_{{\rm F}})}$ and ${\rm maxpool}{(\mathcal{O}_{{\rm F}})}$ into two sets of preliminary weight vectors. The summation of these two weights is then normalized via the Sigmoid function to generate the ultimate channel attention vector $V_{c} \in \mathbb{C}^{1\times 1\times c}$, which is given by
\begin{equation}
	V_{c} = {\rm Sigmoid} \{ {\rm Conv} [  {\rm avgpool}{(\mathcal{O}_{{\rm F}})}  ] + {\rm Conv} [  {\rm maxpool}{(\mathcal{O}_{{\rm F}})}  ]   \}. 
\end{equation} 
By broadcasting and multiplying $V_{c}$ with $\mathcal{O}_{{\rm F}}$, the sub-encoder $E_{3}$ will focus more on the weighted crucial features, thereby optimizing the entire performance.

Overall, assume that the sub-encoder $E_{3}$ has the same structure as $E_{1}$ and $E_{2}$, its outputs can be expressed as
\begin{equation}
	\mathcal{O}_{{\rm E}_{3}} = {\rm BN} ({\rm Relu} ({\rm Conv} (M_{C} \odot \mathcal{O}_{{\rm F}}))) \in \mathcal{Z},
\end{equation}
where $M_{C} \in \mathbb{C}^{h\times w\times c}$ is the channel attention tensor by broadcasting $V_{c}$, with dimensions identical to those of $\mathcal{O}_{{\rm F}}$. As shown in \figref{Corr-MMF module}, $\mathcal{O}_{{\rm E}_{3}}$ represents the data manifold in the latent space $\mathcal{Z}$, which is precisely the output of the Corr-MMF module. For ease of understanding, we denote $\mathcal{O}_{{\rm E}_{3}}$ as $\mathcal{O}_{\rm CorrMMF}$.

Based on the above two-stage encoder, a virtual decoder $D$ is adopted accordingly to further ensure the maximal preservation of key features described in Criterion 1. Note that the decoder $D$ is termed as ``virtual'' because the Corr-MMF module exclusively requires the output $\mathcal{O}_{\rm CorrMMF}$, while $D$ solely serves as an optimization feedback. 

\subsubsection{Correlation Evaluation for Criterion 2} \label{design_c2}
For Criterion 2, we have to develop additional constraints for Corr-MMF module to effectively preserve the correlated information among $\mathcal{G}_{\rm gro}$ and $\mathcal{E}_{h}$. 
For ease of derivation and analysis, denote the two-stage encoder in \figref{Corr-MMF module} as function $f(\cdot)$, where $f(Z) = \mathcal{O}_{\rm CorrMMF}$ with the two-view input $Z = (\mathcal{G}_{\rm gro}, \mathcal{E}_{h})$. Consequently, we can separately obtain the key features from $\mathcal{G}_{\rm gro}$ and $\mathcal{E}_{h}$ by setting $\mathcal{E}_{h}=0$ and $\mathcal{G}_{\rm gro}=0$, denoted as $f(Z_{\mathcal{E}=0})$ and $f(Z_{\mathcal{G}=0})$, respectively. Let $f_{i}(Z_{\mathcal{E}=0})$ and $f_{i}(Z_{\mathcal{G}=0}) $ represent the $i$-th elements of $f(Z_{\mathcal{E}=0})$ and $f(Z_{\mathcal{G}=0})$, respectively, by introducing the adjusted cosine similarity, the correlation can be expressed as
\begin{equation}
	corr[f(Z_{\mathcal{E}=0}), f(Z_{\mathcal{G}=0})] = \dfrac{\sum_{i=1}^{N} \Delta_{{\rm g}, i} \Delta_{{\rm e}, i}}{\sqrt{\sum_{i=1}^{N} \Delta_{{\rm g}, i}^{2}} \sqrt{\sum_{i=1}^{N} \Delta_{{\rm e}, i}^{2}}},
	\label{eq corr}
\end{equation}
where $\Delta_{{\rm g}, i} = f_{i}(Z_{\mathcal{E}=0}) - \overline{f(Z_{\mathcal{E}=0})}$,  $\Delta_{{\rm e}, i} = f_{i}(Z_{\mathcal{G}=0}) - \overline{f(Z_{\mathcal{G}=0})}$,  $\overline{f(Z_{\mathcal{E}=0})}$ and $\overline{f(Z_{\mathcal{G}=0})}$ are the mean values for the key features from $\mathcal{G}_{\rm gro}$ and $\mathcal{E}_{h}$, respectively. Note that maximizing (\ref{eq corr}) empowers the Corr-MMF module to effectively preserve the correlation between features extracted from $\mathcal{G}_{\rm gro}$ and $\mathcal{E}_{h}$, which fulfills the requirements of Criterion 2.

\subsubsection{Objective Function for Corr-MMF Module} \label{la}
Based on the whole network in the above 1) and the correlation evaluation in the above 2), we then develop a matching objective function to train the Corr-MMF module so that it can uniformly satisfy the requirements of Criteria 1 and 2. 

Denote the network in \figref{Corr-MMF module} as $\mathcal{F}$, with its trainable parameters being $\vartheta$. Note that $\mathcal{F}$ is formed by cascading the encoder $f(\cdot)$ with a virtual decoder $D$.
First, we introduce a fusion-reconstruction loss $\mathcal{L}_{\rm fusion}(\vartheta)$ to ensure that the fused features $f(Z)$ can be restored to the original two-view input $Z=(\mathcal{G}_{\rm gro}, \mathcal{E}_{h})$, which is given by
\begin{equation}
	\mathcal{L}_{\rm fusion}(\vartheta) = || \mathcal{F}(Z; \vartheta) - Z ||_{F}.
\end{equation} 
Next, a correlation loss $\mathcal{L}_{\rm corr}(\vartheta)$ is adopted to ensure that the features extracted and fused by Corr-MMF module effectively preserve the correlation between $\mathcal{G}_{\rm gro}$ and $\mathcal{E}_{h}$. According to (\ref{eq corr}), $\mathcal{L}_{\rm corr}(\vartheta)$ is designed as
\begin{equation}
	\mathcal{L}_{\rm corr}(\vartheta) = 1 - corr[f(Z_{\mathcal{E}=0}), f(Z_{\mathcal{G}=0})],
\end{equation}
where $\mathcal{L}_{\rm corr}(\vartheta) \in [0, 2]$. 
Additionally, we also introduce a cross-reconstruction loss $\mathcal{L}_{\rm cross}(\vartheta)$ to assist the process of model training, given by
\begin{equation}
	\mathcal{L}_{\rm cross}(\vartheta) = || \mathcal{F}(Z_{\mathcal{G}=0}; \vartheta) - \mathcal{G}_{\rm gro} ||_{F} + || \mathcal{F}(Z_{\mathcal{E}=0}; \vartheta) - \mathcal{E}_{\rm h} ||_{F}.
\end{equation}
It is worth noting here that the cross-reconstruction loss $\mathcal{L}_{\rm cross}(\vartheta)$ carries the physical meaning in practice: Due to the relativity between $\mathcal{E}_{h}$ and $\mathcal{G}_{\rm gro}$, it is theoretically possible to recover $\mathcal{E}_{h}$ from $\mathcal{G}_{\rm gro}$ and vice versa \cite{jzz}. This process emphasizes not only the reconstruction of $\mathcal{E}_{h}$ and $\mathcal{G}_{\rm gro}$ but also their correlation, serving as a further enhancement for both $\mathcal{L}_{\rm corr}(\vartheta)$ and $\mathcal{L}_{\rm fusion}(\vartheta)$.

Taking all the losses $\mathcal{L}_{\rm fusion}(\vartheta)$, $\mathcal{L}_{\rm corr}(\vartheta)$, and $\mathcal{L}_{\rm cross}(\vartheta)$ into consideration, we present the objective function for Corr-MMF module as follows:
\begin{equation}
	\mathcal{L}_{\rm obj}(\vartheta) = \mathcal{L}_{\rm fusion}(\vartheta) + \mathcal{L}_{\rm cross}(\vartheta) + \lambda \mathcal{L}_{\rm corr}(\vartheta),
\end{equation}
where $\lambda$ is employed to control the proportion among different losses. When $\lambda$ approaches $0$, Corr-MMF module disregards the cross-modal correlation, failing Criterion 2. Conversely, when $\lambda$ tends to infinity, it neglects the preservation of key features and the reconstruction of distinct modalities, violating Criterion 1. Therefore, $\lambda$ requires prudent selection to achieve an optimal balance in Corr-MMF module performance.

\subsection{The MultiModal Representation (MMR) Module}
In conventional multimodal learning tasks, raw multimedia data cannot be directly processed by machines, so it is necessary to conduct the unified description and processing, known as the MultiModal Representation (MMR) \cite{mmf, mmf2}. By definition, MMR refers to the process of representing information from several media in a tensor or vector form \cite{mmf}. While similar to MMF, it places greater emphasis on the uniformity across data representations rather than the feature fusion among different modalities.

In our 3D-CF construction task, the horizontal geographic information $\mathcal{E}_{\rm h}$ and CF measurements $\mathcal{G}_{\rm gro}$ have already been fused by Corr-MMF module ($f(Z) = \mathcal{O}_{\rm CorrMMF}$), hence, we need to further process the vertical geographic information $\mathcal{E}_{v}$ to align with the data representation of $\mathcal{O}_{\rm CorrMMF}$, which motivates our design of the MMR module. 

\begin{figure}[htbp]
	\centering  
	\subfigure[Diagram of the MMR module. It employs an auto-encoder as the core architecture, including an input layer, a downsampling layer, a latent space, an upsampling layer, and an output layer. The spatial attention mechanism (SAM) is introduced as well.]{
			\label{MMR}
			\includegraphics[width=1\linewidth]{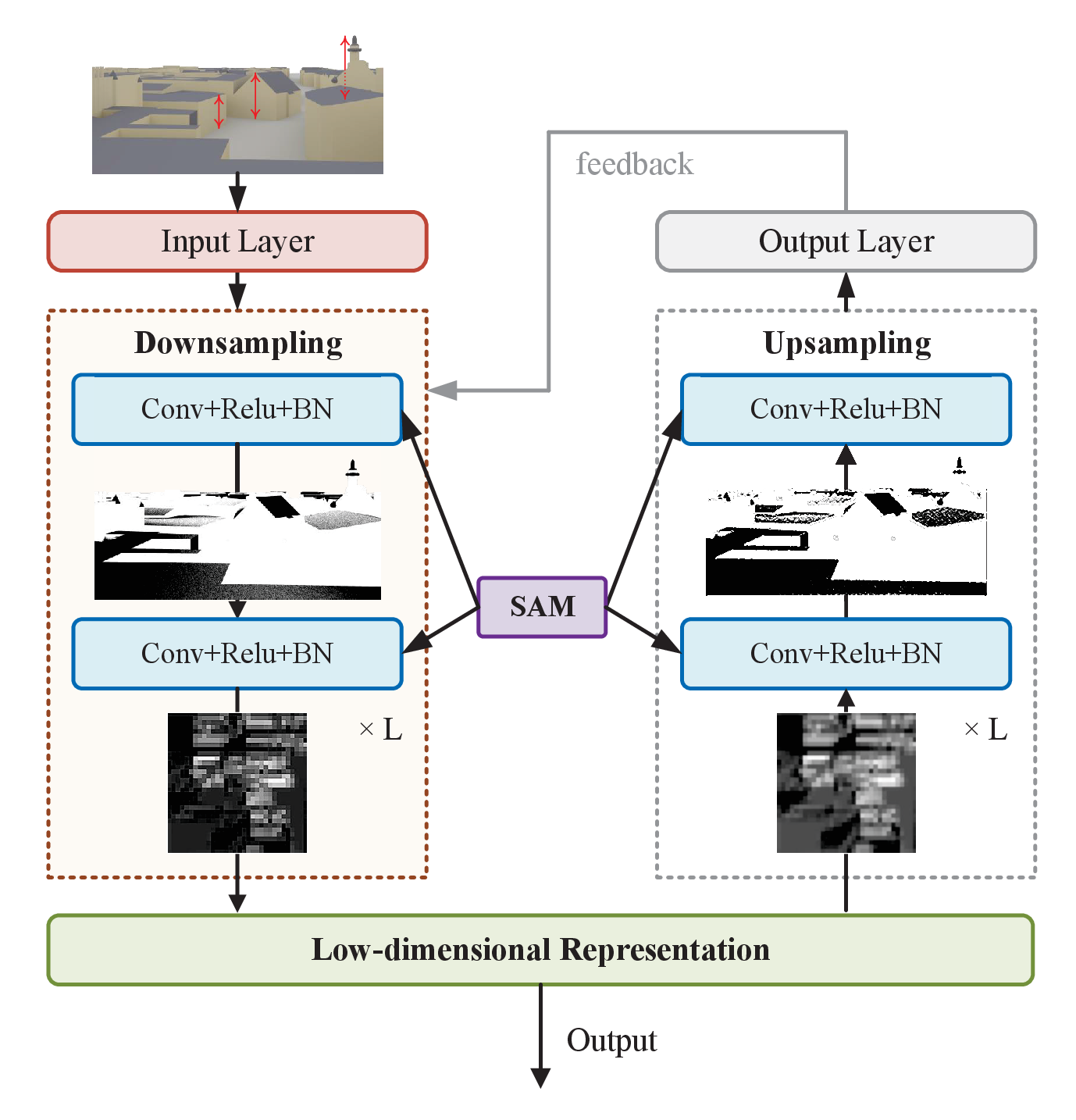}}

	\subfigure[Diagram of the spatial attention mechanism (SAM).]{
			\label{SAM}
			\includegraphics[width=1\linewidth]{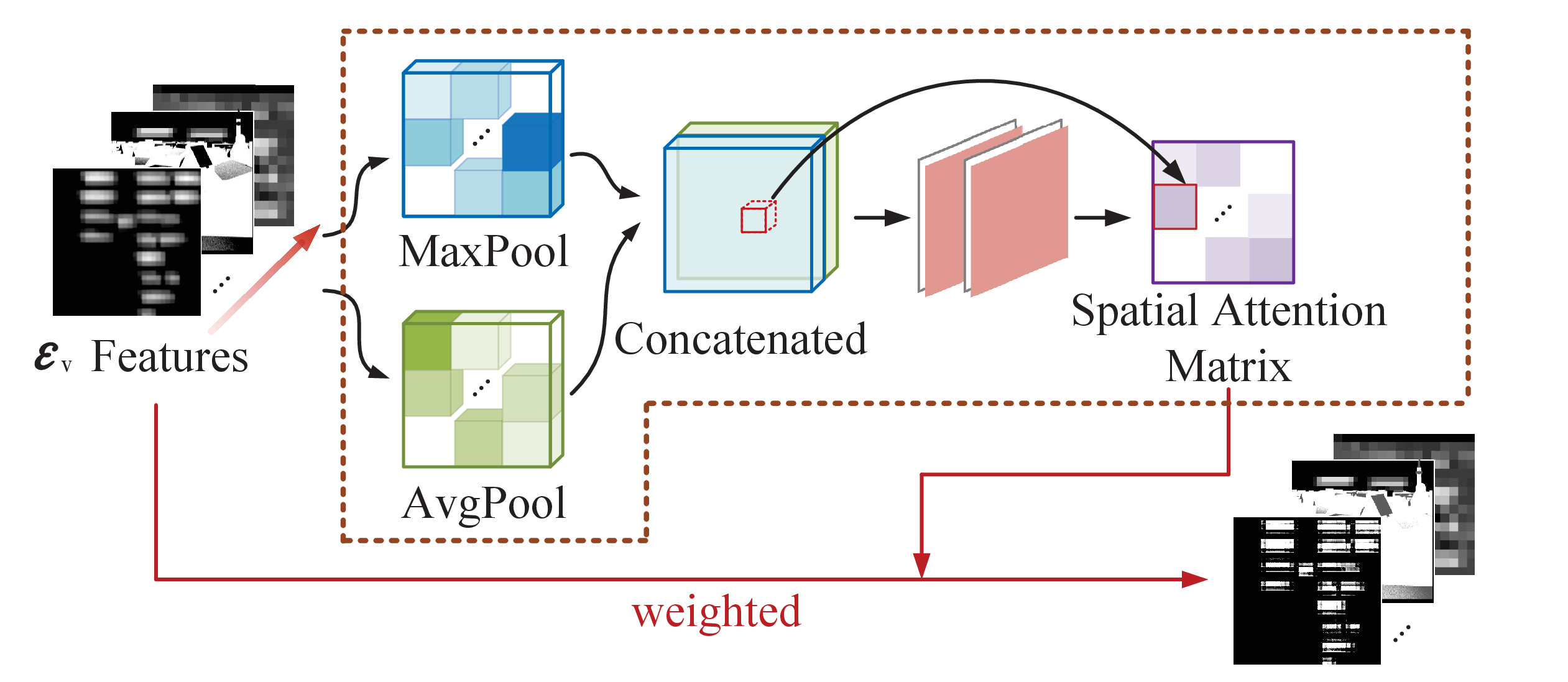}}                       
	\caption{Diagram of the MMR module and its SAM.}
	\label{lossfig}        
\end{figure}

Specifically, the MMR module takes $\mathcal{E}_{v}$ as the input and employs an auto-encoder as its core architecture, which comprises an input layer, a downsampling layer, a latent space, an upsampling layer, and an output layer, as shown in \figref{SAM}. The input layer is primarily utilized for data reshaping, transforming $\mathcal{E}_{v}$ into an easily processable tensor $\mathcal{O}_{\mathcal{E}_{v}} \in \mathbb{C}^{h \times w \times c}$. The following downsampling layer and upsampling layer are similar to the encoder $E_{3}$ and virtual decoder $D$ in \figref{Corr-MMF module}, respectively. It is worth noting here that this structural similarity ensures the uniformity between $\mathcal{O}_{\rm CorrMMF}$ and $\mathcal{O}_{\mathcal{E}_{v}}$, which is the core of the MMR module. The output layer, corresponding to the input layer, is finally appended to reconstruct data back to $\mathcal{E}_{v}$.

Additionally, through a meticulous analysis of the data structure of $\mathcal{O}_{\mathcal{E}_{v}}$, we observe that pixels in feature matrices on each channel imply the specific characteristics of spatial communication environment at different positions. Therefore, the MMR module is expected to adaptively evaluate the importance of each pixel, thus highlighting critical environmental features at key locations. To this end, we introduce a spatial attention mechanism (SAM) to assign varying weights to different pixels in each channel of $\mathcal{O}_{\mathcal{E}_{v}}$ \cite{cbam}, as depicted in \figref{SAM}. Specifically, SAM employs a global max pooling and a global average pooling to separately obtain feature maps of each pixel fibers in $\mathcal{O}_{\mathcal{E}_{v}}$, denoted as ${\rm maxpool}(\mathcal{O}_{\mathcal{E}_{v}}) \in \mathbb{C}^{h\times w\times 1}$ and ${\rm avgpool}(\mathcal{O}_{\mathcal{E}_{v}}) \in \mathbb{C}^{h\times w\times 1}$, respectively. Subsequently, we construct the integrated statistical information by concatenating ${\rm maxpool}(\mathcal{O}_{\mathcal{E}_{v}})$ and ${\rm avgpool}(\mathcal{O}_{\mathcal{E}_{v}})$, and feed it into a convolutional layer to derive the preliminary weight matrix. Afterwards, the Sigmoid function is introduced for normalization to acquire the ultimate spatial attention matrix $V_{s}\in\mathbb{C}^{h\times w\times 1}$, which is given by
\begin{align}
	&V_{s} \nonumber\\
	&={\rm Sigmoid}\{{\rm Conv}[ {\rm Concat}( {\rm avgpool}(\mathcal{O}_{\mathcal{E}_{v}}), {\rm maxpool}(\mathcal{O}_{\mathcal{E}_{v}})) ]\}.
\end{align}
By broadcasting and multiplying $V_{s}$ with $\mathcal{O}_{\mathcal{E}_{v}}$, the downsampling process will focus more on the weighted crucial environmental features at key locations, thereby improving the performance of the MMR module. 

Overall, the output of the MMR module, namely the low-dimensional representation of $\mathcal{E}_{v}$, can be expressed as
\begin{equation}
	\mathcal{O}_{\rm MMR} = {\rm BN}( {\rm Relu}( {\rm Conv}(M_{S}\odot \mathcal{O}_{\mathcal{E}_{v}} ) ) ),
\end{equation}
where $M_{S} \in \mathbb{C}^{h\times w\times c}$ is the spatial attention tensor by broadcasting $V_{s}$, with dimensions identical to those of $\mathcal{O}_{\mathcal{E}_{v}}$.

\subsection{The Channel State Information Regression (CSI-R) Module}
In the 3D-CF construction task, $\mathcal{E}$ and $\mathcal{G}_{\rm gro}$ have been, respectively, transformed into low-dimensional representations $\mathcal{O}_{\rm CorrMMF}$ and $\mathcal{O}_{\rm MMR}$ with identical data structures, where $\mathcal{O}_{\rm CorrMMF}$ embodies the 3D-CF patterns in the horizontal direction and $\mathcal{O}_{\rm MMR}$ embodies the environment patterns in the vertical direction. Based on this, we next design the CSI-R module to predict the channel information at a specific LAV location $\mathcal{X}$ in low-altitude airspace, thereby enabling the flexible 3D-CF construction free from grid constraints.

\begin{figure}[htbp]
	\centering
	\includegraphics[width=1\linewidth]{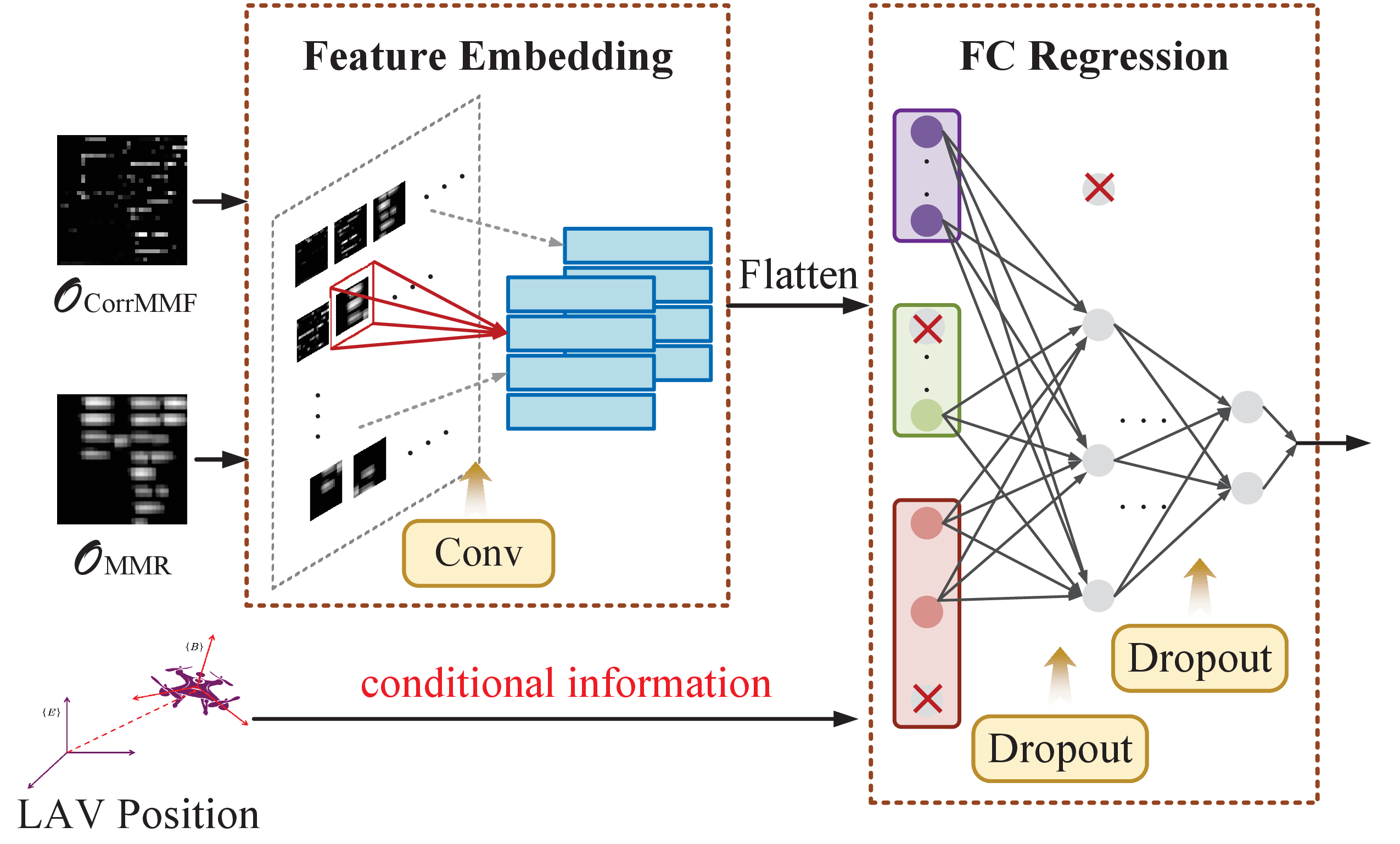}
	\caption{Diagram of the proposed CSI-R module. The network includes a feature embedding layer and a fully connected regression layer.}
	\label{CSIR}
\end{figure}

As illustrated in \figref{CSIR}, The CSI-R module uses $\mathcal{O}_{\rm CorrMMF}$ and $\mathcal{O}_{\rm MMR}$ as inputs and consists of a feature embedding layer and a fully connected (FC) regression layer. Regarding the feature embedding layer, we first partition the feature maps of $\mathcal{O}_{\rm CorrMMF}$ and $\mathcal{O}_{\rm MMR}$ into several equally-sized patches by a sliding window. Then, a convolutional layer with its kernel size matching the patch size is followed to generate a sequence of embedding vectors. Flattening these sequences yields a one-dimensional embedding output, which will be fed into the subsequent FC regression layer. Note that the entire feature embedding process captures elemental similarities in the feature maps of $\mathcal{O}_{\rm CorrMMF}$ and $\mathcal{O}_{\rm MMR}$, respectively, positioning elements with higher similarity closer in the embedding space, thus accelerating the convergence and enhance the accuracy. The FC regression consists of several dense layers, each followed by a dropout layer to prevent severe overfitting. During the regression process, 3D coordinates of the LAV are vectorized and incorporated as conditional information, prompting the prediction of location-specific channel characteristics for 3D-CF construction.

In summary, the mapping relationship $\Psi$ can be effectively approximated through the cascade of Corr-MMF module, MMR module, and CSI-R module, thereby enabling the construction of CSI-tuples in (\ref{3dcf}) and achieving the configuration of 3D-CF according to practical communication requirements. This certainly provides novel insights into the low-altitude communications.

\section{Numerical Results} \label{S3}
In this section, numerical results are provided to evaluate the performance of the proposed 3D-CF multimodal framework. First, we introduce the generation of our Sionna-based datasets and detail the experiment setup. Then, we explore the impact of $\lambda$ on the accuracy of 3D-CF construction. By employing Kriging interpolation, GPR, GAN, and FL as benchmarks, we next compare the 3D-CF construction performance across different scenarios, demonstrating the prediction accuracy and generalization capability of the proposed 3D-CF multimodal framework. Finally, we analyze the computational complexity.

\subsection{Datasets}
All datasets are generated by NVIDIA Sionna, an open-source library for research on wireless communication systems \cite{sionna}. Specifically, we first download the geographic environment maps of Nanjing, China, from OpenStreetMap (OSM) \cite{osm}, a collaborative mapping project maintained by a global community of volunteers who continuously update and validate geographic information. These maps are then imported into Blender to construct communication scenarios, which are subsequently transferred to Sionna for dataset generation via ray tracing (RT). Note that all selected areas represent typical urban macro-cell or micro-cell scenarios with varying building shapes, quantities, and distributions, thereby ensuring strong data diversity to support model training and facilitate the evaluation of generalization capability.

To facilitate our experiments, the BS is randomly deployed in the target region with $N_{\rm BS}= 64$, and the LAV flight altitudes are confined to $25-80$ m. CF measurement data $\mathcal{G}_{\rm gro}$ are uniformly sampled at a height of $1.5$ m with a resolution of $1$ m. Note that there exists an inherent resolution-accuracy-complexity trade-off: enhanced near-ground sampling density improves the accuracy of 3D-CF construction, but increases the computational complexity. Hence, the resolution of $\mathcal{G}_{\rm gro}$ should be determined based on practical limitations in real-world scenarios. Based on the specific interface for wireless communication simulation in Sionna, we conduct ray tracing to obtain all possible LOS and NLOS paths between BS and LAVs, establish the channel model in accordance with (\ref{cmo}), compute RSS according to (\ref{gm}), and construct 3D-CF by (\ref{3dcf}). 
To further enhance the stability of network training, we adopt the ``max-min'' linear normalization to scale the raw RSS in 3D-CF into $[0,1]$, which is given by
\begin{equation}
	g_{m}' = \max \left\lbrace \dfrac{g_{m}-g_{\rm thr}}{(g_{m})_{\rm max}-g_{\rm thr}}, 0  \right\rbrace,
\end{equation}
where $g_{\rm thr}$ is the RSS threshold since signals below $g_{\rm thr}$ are practically undetectable by LAV in real-world scenarios \cite{2den3}.
Additional parameters used to generate the datasets are displayed in \tabref{tb:data parameters}.

\begin{table}[htbp]
	\captionsetup{font=footnotesize}
	\captionsetup{justification=centering}
	\caption{Parameters used to generate datasets in the platform of Sionna.}
	\label{tb:data parameters}
	\centering
	\ra{1.5}
	\scriptsize
	\begin{tabular}{LR}
		\toprule
		Parameters                                         &   Value \\
		\midrule
		\rowcolor{lightblue}
		Communication scenario                   & urban macro-cell / micro-cell \\
		Size of the target region                    & $256\times 256$  \\
		\rowcolor{lightblue}
		Number of antenna elements for BS  & $64$ \\
		BS height                                          &  $25$ m \\		
		\rowcolor{lightblue}
		LAV flight altitudes                           & $25-80$ m \\
		Sampling height of $\mathcal{G}_{\rm gro}$       & $1.5$ m \\
		\rowcolor{lightblue}
		Carrier frequency                              & $3.5$ GHz \\
		Maximum number of paths               & $5$ \\
		\rowcolor{lightblue}
		RSS threshold                                    & $-147$ dB \\
		Transmit power                                & $23$ dBm \\
		\bottomrule
	\end{tabular}
\end{table}

\subsection{Experiment Setup}
In correspondence with the 3D-CF multimodal framework in Section \ref{framework section}, we configure all hyper-parameters of the three modules in \tabref{hyper parameters}. Specifically, regarding the Corr-MMF module, the batch size is set to be $128$ and the epochs are $60$, with the learning rate programmed to be $0.0001$ for the first $35$ epochs and then linearly decaying to zero. For the MMR module, the training epochs are set to be $60$, and the learning rate is initialized at $0.001$ for the first $35$ epochs and subsequently reduced linearly to zero. The training process of the CSI-R module requires $15$ epochs with a reduced batch size of $32$. The learning rate is set to be $0.0001$ for the first $10$ epochs, followed by a linear decay to zero. All the simulations are implemented by TensorFlow, with the computer equipped with an Intel(R) Core(TM) i7-12700 and a GeForce GTX 4090.

To evaluate the performance of the proposed multimodal framework in 3D-CF construction, we employ the mean absolute error (MAE) and root mean square error (RMSE) as metrics, which can be expressed as
\begin{align}
	{\rm MAE} & = \dfrac{1}{n} \sum_{i=1}^{n} |\tilde{g}_{m}' - g_{m}'|,
\end{align}

\begin{align}
	{\rm RMSE} & = \sqrt{ \dfrac{1}{n} \sum_{i=1}^{n} (\tilde{g}_{m}' - g_{m}')^{2}},
\end{align}
where $\tilde{g}_{m}'$ is the predicted RSS in CSI-tuples in 3D-CF. Since MAE provides equal weights to all errors, it can intuitively reflect the overall 3D-CF construction performance. Conversely, RMSE exhibits greater sensitivity to outliers, facilitating the detection of extreme prediction deviations in our proposed 3D-CF multimodal framework. The combination of these two metrics enables a more comprehensive evaluation of the experimental results.   

\begin{table}[htbp]
	\captionsetup{font=footnotesize}
	\captionsetup{justification=centering}
	\caption{Hyper-parameters of Corr-MMF, MMR, and CSI-R modules in 3D-CF MultiModal framework.}
	\label{hyper parameters}
	\centering
	\ra{1.5}
	\scriptsize
	\begin{tabular}{LccR}
		\toprule
		\multirow{2}{*}{Parameter}  &  \multicolumn{3}{c}{Module} \\
		\cline{2-4}
		& \ \ \ \ Corr-MMF \ \ \ \   & \ \ \ \ MMR \ \ \ \  & \ \ \ \ CSI-R \ \ \ \ \\
		\midrule
		\rowcolor{lightblue}
		Epochs                               & 60               & 50        & 15 \\
		Delay Epochs                     & 35               & 35        & 10 \\
		\rowcolor{lightblue}
		Learning Rate                    & 0.005          &  0.001    & 0.0001 \\
		Batch Size                          & 128            & 128        & 32  \\
		\rowcolor{lightblue}
		Optimizer                          & \multicolumn{3}{c}{Adam} \\
		\bottomrule
	\end{tabular}
\end{table}

\subsection{Influence of $\lambda$ in the Corr-MMF module}
In Corr-MMF module, $\lambda$ would influence the network performance by controlling the proportions among different losses. Therefore, in this section, we examine its impact on the 3D-CF construction error and analyze its effect on the convergence behavior of the Corr-MMF module.

\begin{figure}[htbp]
	\centering
	\includegraphics[width=1\linewidth]{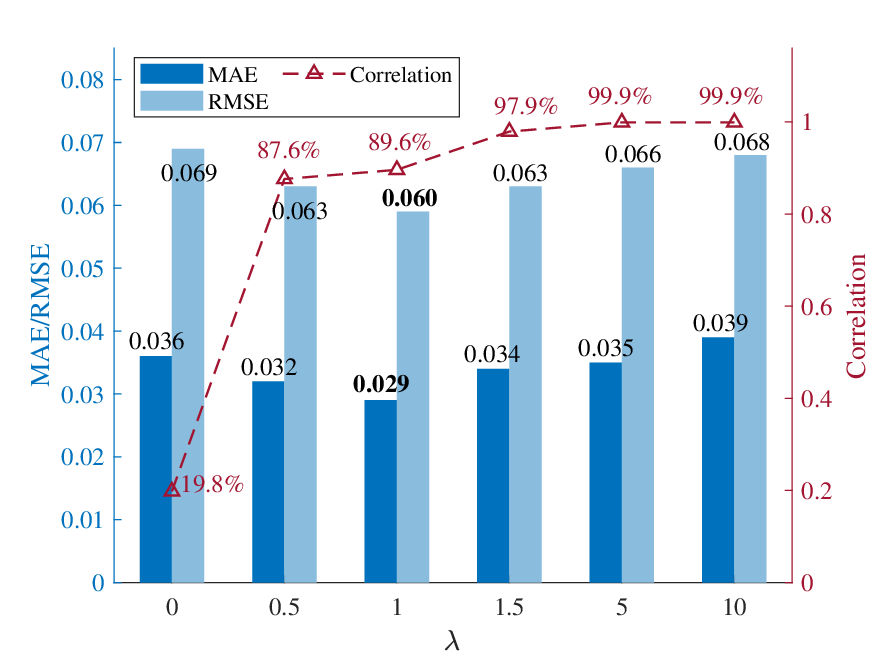}
	\caption{MAE and RMSE performance of the 3D-CF multimodal framework under different $\lambda$.}
	\label{tb:lamda}
\end{figure}

As shown in \figref{tb:lamda}, the proposed multimodal framework can achieve the minimal 3D-CF construction error at $\lambda=1$, with ${\rm MAE}=0.029$ and ${\rm RMSE}=0.060$, capturing approximately $89.6$\% of the correlation between $\mathcal{E}_{h}$ and $\mathcal{G}_{\rm gro}$.
As $\lambda$ gradually decreases, the weight of $\mathcal{L}_{\rm corr}(\vartheta)$ reduces and the 3D-CF construction error increases progressively. Notably at $\lambda=0$, the performance of MAE and RMSE deteriorates by $24.1$\% and $15$\% respectively, with captured correlation between $\mathcal{E}_{h}$ and $\mathcal{G}_{\rm gro}$ dropping to $19.8$\%. In practice, an excessively small $\lambda$ will prevent the Corr-MMF module from extracting correlated features between $\mathcal{E}_{h}$ and $\mathcal{G}_{\rm gro}$, thereby disabling the operation of feature fusion and degrading the construction performance of the proposed 3D-CF multimodal framework.
Similarly, as $\lambda$ increases, the weight of $\mathcal{L}_{\rm corr}(\vartheta)$ rises and the performance of 3D-CF construction declines. Especially at $\lambda=10$, where $\lambda\mathcal{L}_{\rm corr}(\vartheta)$ can be considered infinite compared with the magnitude of other losses, the performance of MAE and RMSE deteriorates by $34.4$\% and $13.3$\%, respectively. Here, an excessively large $\lambda$ will unreasonably overemphasize the correlation between $\mathcal{E}_{h}$ and $\mathcal{G}_{\rm gro}$ (capturing as much as 99.9\%), completely suppressing their distinctive features and leading to a collapse of the Corr-MMF module.
To conclude, the determination of $\lambda$ must strike an effective balance between distinctive feature extraction and correlated feature fusion, thus optimizing the performance of Corr-MMF module and achieving accurate 3D-CF construction.

\begin{figure}[htbp]
	\centering
	\includegraphics[width=1\linewidth]{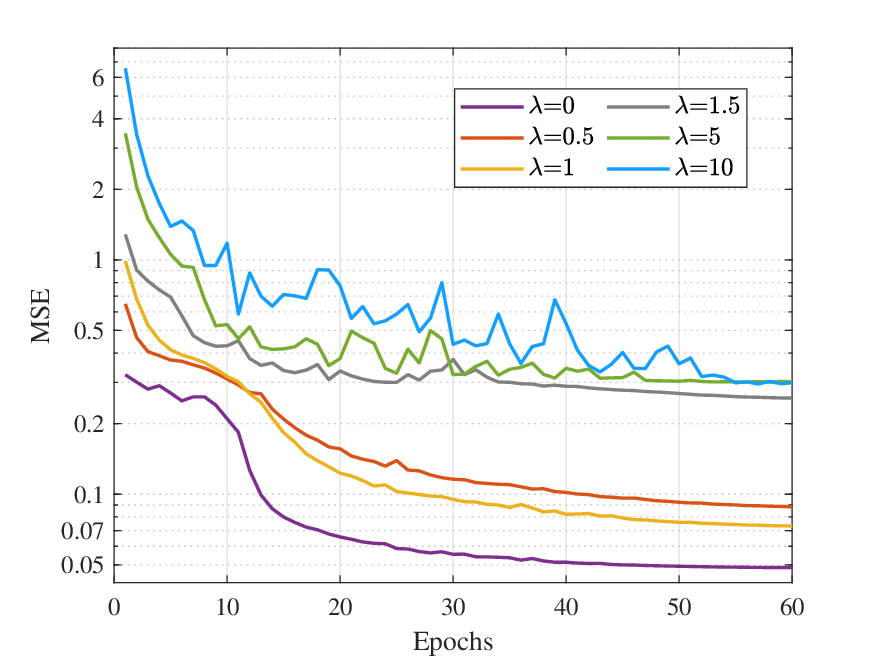}
	\caption{The convergence behavior of the validation loss for the Corr-MMF module under different $\lambda$.}
	\label{loss}
\end{figure}

\figref{loss} presents the convergence behavior of the validation loss for the Corr-MMF module under different $\lambda$. A fundamental observation is that larger $\lambda$ corresponds to greater $\lambda\mathcal{L}_{\rm corr}(\vartheta)$, consequently resulting in higher values of the objective function $\mathcal{L}_{\rm obj}(\vartheta)$. Nevertheless, the validation loss is lower when $\lambda=1$ compared to $\lambda=0.5$, which demonstrates that $\lambda=1$ can optimize the network performance to the greatest extent. This result is consistent with the conclusion obtained in \figref{tb:lamda}.
Furthermore, smaller $\lambda$ leads to smoother loss curves and contributes to a more stable network training, whereas larger $\lambda$ results in greater fluctuations. This occurs because smaller $\lambda$ balances the magnitudes of different losses, enabling steadier execution of the gradient descent algorithm. Conversely, larger $\lambda$ causes $\mathcal{L}_{\rm corr}(\vartheta)$ to be more dominant, making the network significantly more susceptible to its oscillation and hence resulting in fluctuations across different epochs.
To conclude, variations in $\lambda$ substantially impact the stability of network training.

\subsection{Evaluation of the 3D-CF Construction Performance}
In this section, we compare the performance of the proposed 3D-CF multimodal framework with four benchmarks and evaluate the generalization capability under different scenarios.

\subsubsection{Benchmarks}
To evaluate the performance of our proposed 3D-CF multimodal framework, the Kriging interpolation \cite{kriging}, GPR \cite{GPR}, GAN \cite{GAN1}, and FL \cite{MLP} are adopted as benchmarks.
\begin{itemize}
	\item Kriging interpolation \cite{kriging}:	Kriging is a classical interpolation method for constructing 3D-CF. Based on the prior sampling data, it achieves RSS estimation at arbitrary locations by incorporating distances and modeling 3D spatial correlation through different variograms. The Kriging interpolation method does not require a grid-based model for CF, thus making the flexible 3D-CF construction possible.
	\item GPR \cite{GPR}: Gaussian process regression is a widely used statistical non-parametric model for 3D-CF construction. It can construct the optimal approximator of RSS distribution by designing specific kernel functions based on signal propagation characteristics. This GPR-based method does not require the grid-based model as well, therefore making the 3D-CF construction more flexible. 
	\item GAN \cite{GAN1}: Generative adversarial network is a significant deep generative model employed for 3D-CF construction. Unlike the Kriging-based method and GPR-based method, GAN perceives the communication environment and utilizes it as conditional information to generate 3D-CF. However, it requires the grid-based model to represent CF as a multi-channel image, which necessitates a uniform partitioning of the target area and assigns the same RSS for all LAVs within the same grid, resulting in lower accuracy and reduced flexibility.
	\item FL \cite{MLP}: Federated learning represents a state-of the-art framework to recover 3D-CF. Embedded with deep neural networks, it enables the collaborative utilization of data from multiple LAVs to establish the global 3D-CF. Since the regression-based FL framework requires no special assumptions about the CF model, it balances both accuracy and flexibility in the process of 3D-CF construction.
\end{itemize}
The sampling rate for Kriging interpolation and GPR is set to be $5$\%, while the training, validation, and test datasets are identical for all other ML-based methods. 

\begin{figure*}[t]
	\centering  
	\subfigure[3D-CF for scenario 1.]{
			\includegraphics[width=0.235\linewidth]{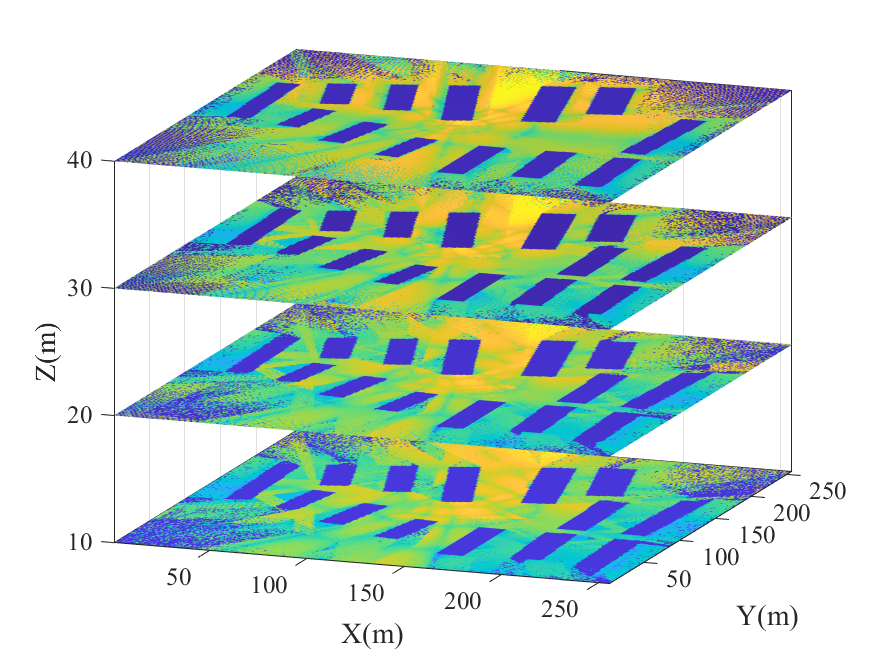}}
	\subfigure[3D-CF for scenario 2.]{
			\includegraphics[width=0.235\linewidth]{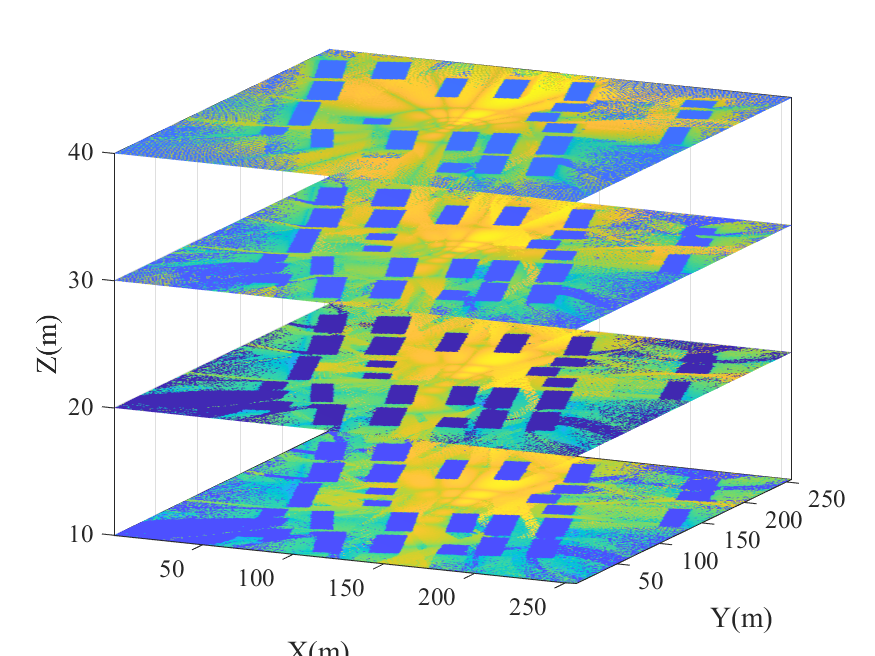}} 
	\subfigure[3D-CF for scenario 3.]{
			\includegraphics[width=0.235\linewidth]{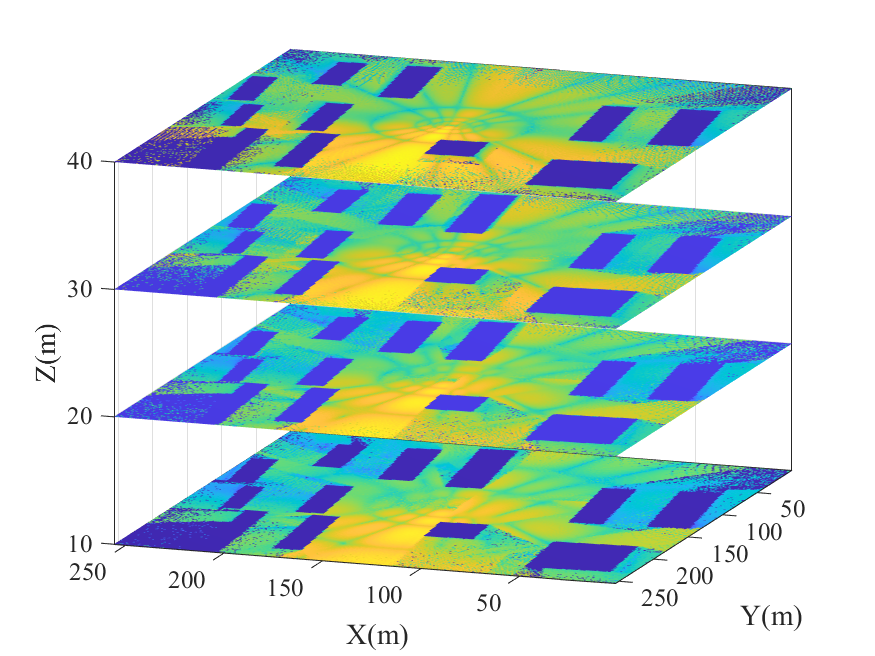}}    
	\subfigure[3D-CF for scenario 4.]{ 
		   \includegraphics[width=0.235\linewidth]{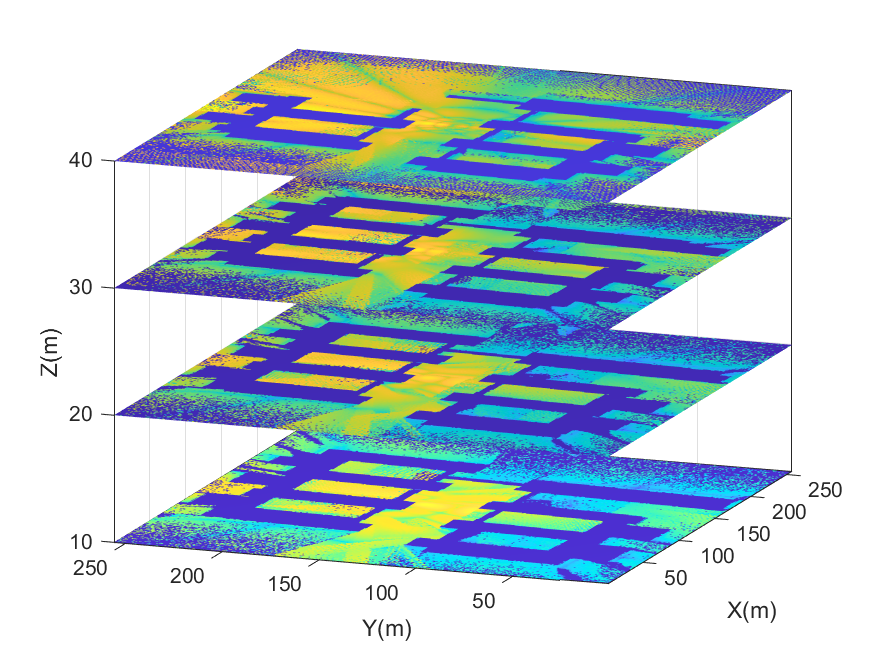}} 		                   
	\caption{Illustrations of 3D-CF under four randomly selected communication scenarios. The RSS distribution at four horizontal planes, specifically at heights of 10 m, 20 m, 30 m, and 40 m, are illustrated along the Z axis.}      
	\label{scenarios}
\end{figure*}

\begin{table*}[htbp]
	\captionsetup{font=footnotesize}
	\captionsetup{justification=centering}
	\caption{Comparison of the 3D-CF multimodal framework with state-of-the-art approaches under different communication scenarios regarding the RMSE and MAE performance.}
	\label{tb:scene}
	\centering
	\ra{1.5}
	\scriptsize
	\begin{tabular}{LcccccccR}
		\toprule
		\multirow{2}{*}{ } & \multicolumn{2}{c}{Scenario 1} & \multicolumn{2}{c}{Scenario 2} & \multicolumn{2}{c}{Scenario 3} & \multicolumn{2}{c}{Scenario 4}\\
		\cline{2-9}
	                  & \ \ RMSE \ \ &  \ \ MAE  \ \ & \ \ RMSE  \ \ & \ \ MAE  \ \ &  \ \ RMSE  \ \ &  \ \ MAE  \ \ &  \ \ RMSE  \ \ &  \ \ MAE  \ \ \\
		\midrule
		\rowcolor{lightblue}
        Kriging \cite{kriging} & 0.522 & 0.467 & 0.348 & 0.287 & 0.340 & 0.289 & 0.459 & 0.384 \\
        GPR \cite{GPR}    & 0.343 & 0.223 & 0.291 & 0.201 & 0.304 & 0.161 & 0.325 & 0.194 \\  
        \rowcolor{lightblue}
        GAN \cite{GAN1}   & 0.234 & 0.195 & 0.208 & 0.156 & 0.269 & 0.221 & 0.266 & 0.223 \\
        Federated Learning \cite{MLP} & 0.088 & 0.067 & 0.041 & 0.034 & 0.035 & 0.029 & 0.056 & 0.039 \\
        \rowcolor{lightblue}
        Multimodal framework (ours) & \textbf{0.069}$\downarrow$ & \textbf{0.044} $\downarrow$ & \textbf{0.026} $\downarrow$ & \textbf{0.019} $\downarrow$ & \textbf{0.024} $\downarrow$ & \textbf{0.021} $\downarrow$ & \textbf{0.045} $\downarrow$ & \textbf{0.027} $\downarrow$\\
		\bottomrule
	\end{tabular}
\end{table*}

\subsubsection{Comparison to Different Benchmarks}
\tabref{tb:scene} presents the comparison of 3D-CF construction performance between the proposed multimodal framework and the four other baselines.
Compared to the non-AI methods like Kriging interpolation \cite{kriging} and GPR \cite{GPR}, the 3D-CF multimodal framework demonstrates a reduction in RMSE by factors of $7.5$ and $4.9$, respectively, and a decrease in MAE by factors of $10.6$ and $5.1$, respectively, significantly enhancing the accuracy of 3D-CF construction. Fundamentally, the Kriging-based method and GPR merely fit the data itself without exploring the impact of communication environments on channel characteristics. Particularly in low-altitude airspace, where the physical environments become more complex, pure data fitting is no longer sufficient to accurately predict the distribution of channel information.
Secondly, compared to the classical generative model GAN \cite{GAN1}, the proposed multimodal framework can achieve $3.4$-fold and $4.4$-fold performance advantages in RMSE and MAE metrics, respectively. One point should be noted that in low-altitude airspace, the physical environment comprises different data modalities in horizontal and vertical dimensions, with LAV coordinates evolving into ternary arrays. Since GAN in \cite{GAN1} is limited to processing merely single-modal data, the 3D-CF construction accuracy is inevitably compromised. Moreover, the adoption of GAN needs to model 3D-CF as images, resulting in a rigid and inflexible construction and cannot achieve the non-uniform density adaptation according to practical communication requirements. Our proposed multimodal framework effectively addresses these limitations, thereby significantly reducing 3D-CF construction errors and enhancing flexibility.
Finally, compared to the state-of-the-art method, the proposed multimodal framework still outperforms the FL-based approach \cite{MLP} by $27.5$\% and $52.2$\% in terms of the RMSE and MAE, respectively. This stems from the operation of feature extraction and feature fusion for diverse data modalities via Corr-MMF module and MMR module, which enables the final regression network to comprehensively learn the mapping relationship between LAV's positions and its RSS.
To conclude, owing to the CSI-tuples-based model and the module-based design, the proposed multimodal framework exhibits superior performance and high flexibility in 3D-CF construction.

\subsubsection{Comparison Under Different Scenarios}
\figref{scenarios} presents the 3D-CF across four randomly selected communication scenarios with different urban structures or building densities. For illustrative purposes only, the RSS distribution across four horizontal planes at 10 m, 20 m, 30 m, and 40 m is given along the Z‑axis. \tabref{tb:scene} provides the comparison of RMSE and MAE performance across these scenarios. It can be observed that, regardless of the scenario, our proposed multimodal framework can always achieve 3D-CF construction with smaller errors than the baselines, demonstrating its generalization ability with RMSE and MAE standard deviations of $0.0012$ and $0.0014$, respectively.

\subsection{Ablation Experiments}
In this subsection, we conduct two ablation experiments to respectively demonstrate the contributions of different input modalities and attention mechanisms to the proposed 3D-CF multimodal framework.

\subsubsection{Input Modality}
\tabref{miss-modal} presents the contribution of each input modality to the proposed 3D-CF multimodal framework. Firstly, when $\mathcal{E}_{v}$ is missing, the model fails to learn the vertical signal-propagation characteristics, leading to a rapid performance deterioration. This outcome indicates that the construction of 3D-CF must account for the three-dimensional nature of the low-altitude environment. If existing 2D-CF construction methods are directly applied to low-altitude scenarios that only consider the horizontal building distribution, severe model mismatch will arise. Secondly, when both $\mathcal{E}_{h}$ and $\mathcal{G}_{\rm gro}$ are missing, the model cannot learn the horizontal characteristics of the CSI distribution, leading to significant degradation in model performance as well. However, when only one of them is missing, the 3D-CF constructed by the proposed architecture experiences only a 4.4\% drop in performance. In practice, both horizontal environmental information and near-ground measurements can reflect the signal-propagation characteristics in the horizontal direction, and the absence of either alone does not cause model collapse.

\begin{table}[htbp]
	\captionsetup{font=footnotesize}
	\captionsetup{justification=centering}
	\caption{Contribution of each input modality to the 3D-CF multimodal framework.}
	\label{miss-modal}
	\centering
	\ra{1.5}
	\scriptsize
	\begin{tabular}{ccc|cc}
		\toprule
		\multicolumn{2}{c}{Geographic environments $\mathcal{E}$} & \multirow{2}{*}{ Measurements $\mathcal{G}_{\rm gro}$ } & \multicolumn{2}{c}{Evaluation metrics} \\
		\cline{1-2} \cline{4-5}
		$\mathcal{E}_{h}$ & $\mathcal{E}_{v}$ &  & RMSE & MAE \\
		\midrule
		\rowcolor{lightblue}
		\Checkmark          & \Checkmark          & \Checkmark          & \textbf{0.045} & \textbf{0.027} \\
		\Checkmark          & \Checkmark          & \XSolidBrush        & 0.047  & 0.028 \\
		\rowcolor{lightblue}
		\XSolidBrush        & \Checkmark            & \Checkmark          & 0.047 & 0.029 \\
		\Checkmark          & \XSolidBrush           & \Checkmark           & 0.687 & 0.675 \\
		\rowcolor{lightblue}
		\XSolidBrush        & \XSolidBrush           & \Checkmark          & 0.700 & 0.692 \\
		\bottomrule
	\end{tabular}
\end{table}

In summary, the ablation study on input modalities demonstrates the effectiveness and necessity of each type of prior information, thereby validating the rationality of our proposed 3D-CF multimodal framework.

\subsubsection{Attention Mechanisms}
\tabref{ablation experiments} presents the impact of three different attention mechanisms, including TAM, CAM, and SAM, on the 3D-CF construction. Firstly, the model performance deteriorates severely when the TAM is removed, whereas the degradation is less pronounced when the CAM or SAM is removed. In principle, the TAM operates directly on the inputs, primarily strengthening the near-ground measurement data near the LAV projection. In contrast, the CAM and SAM function internally within the network, emphasizing key features through self-learning. Once the TAM is absent, the data itself lacks essential weighting, leading to severe performance degradation regardless of the internal network design.
Secondly, the performance decline in the absence of CAM is less severe than that when the SAM is missing. Since both the CAM and TAM are mechanisms within the Corr-MMF module, even if CAM is removed, the TAM mechanism still enables this module to achieve a relatively effective training result.

\begin{table}[htbp]
	\captionsetup{font=footnotesize}
	\captionsetup{justification=centering}
	\caption{Ablation study on different attention mechanisms for 3D-CF construction.}
	\label{ablation experiments}
	\centering
	\ra{1.5}
	\scriptsize
	\begin{tabular}{ccc|cc}
		\toprule
		\multicolumn{2}{c}{\ \ Corr-MMF module \ \ } & \ \  MMR module \ \  & \multicolumn{2}{c}{ \ \  Evaluation metrics \ \ } \\
		\cline{1-3} \cline{4-5}
		CAM & TAM & SAM & RMSE & MAE \\
		\midrule
		\rowcolor{lightblue}
		\Checkmark          & \Checkmark          & \Checkmark          & \textbf{0.045} & \textbf{0.027} \\
		\XSolidBrush          & \Checkmark          & \Checkmark        & 0.065  & 0.052 \\
		\rowcolor{lightblue}
		\Checkmark        & \XSolidBrush            & \Checkmark          & 0.546 & 0.455 \\
		\Checkmark          &  \Checkmark          &  \XSolidBrush          & 0.290 & 0.282 \\
		\bottomrule
	\end{tabular}
\end{table}

In summary, the ablation study on attention mechanisms demonstrates that all TAM, CAM, and SAM play indispensable roles in the proposed multimodal framework, enabling the accurate and efficient 3D-CF construction.

\subsection{Complexity Comparison}
In this subsection, we analyze the complexity of the proposed 3D-CF multimodal framework versus benchmarks and present comparative results of their inference times. 

\begin{figure}[htbp]
	\centering
	\includegraphics[width=1\linewidth]{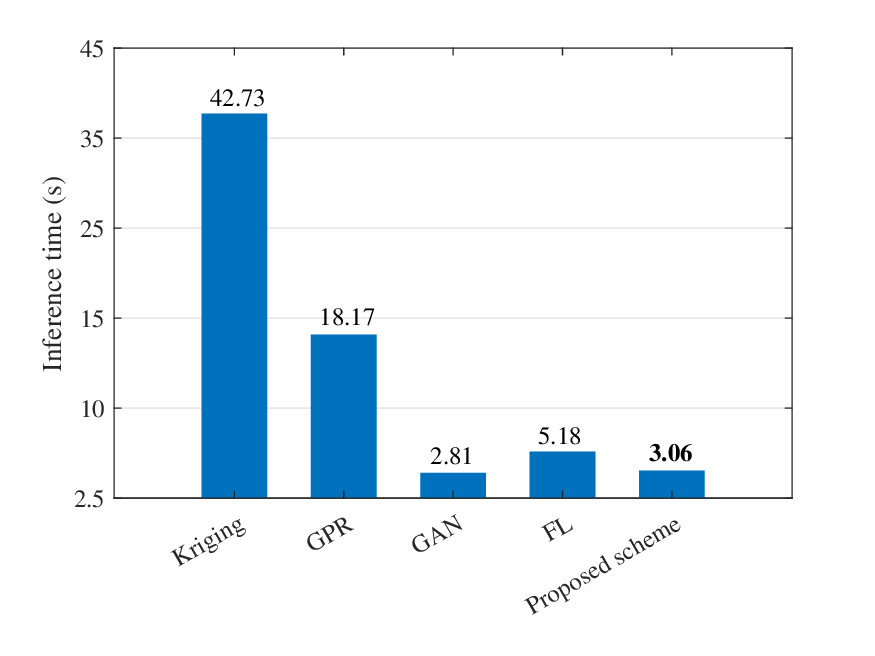}
	\caption{Comparison of the 3D-CF inference time between the proposed multimodal framework and state-of-the-art approaches.}
	\label{time}
\end{figure}

Define $N$ as the amount of training data.
For the Kriging-based method, we must solve the Kriging system of equations with a size of $N\times N$ to complete the construction of 3D-CF, the complexity of which remains $\mathcal{O}(N^{3})$ even when employing the Cholesky decomposition \cite{krigingcom}.
For the GPR-based method, the complexity of covariance matrix inversion is $\mathcal{O}(N^{2})$, and the complexity of marginal likelihood approximation in the process of solving posterior probability is $\mathcal{O}(N)$. Consequently, the overall complexity for the GPR-based 3D-CF construction is $\mathcal{O}(N^{3})$ \cite{gprcom}. 
For the GAN-based approach, its complexity can be expressed as $ \sum_{\ell=1}^{L_{\rm GAN}} \mathcal{O}_{\ell}(N C_{\ell-1} H_{\ell}W_{\ell}C_{\ell} K^{2}_{\ell}) $, where $L_{\rm GAN}$ is the number of convolution layers, $ K_{\ell} $ is size of the convolution kernel, and $ H_{\ell}, W_{\ell}$, and $ C_{\ell} $ are heights, widths, and channels of the $ \ell $-th layer's output, respectively. 
For the FL-based approach which is mainly composed of the multilayer perceptron, its complexity is given by $\sum_{p=1}^{P_{\rm FL}}\mathcal{O}(N d_{p} h_{p})$, where $P_{\rm FL}$ is the number of hidden layers, $d_{p}$ is the input dimension and $h_{p}$ represents the number of neurons in the $p$-th hidden layer.
Regarding our proposed 3D-CF multimodal framework, its complexity is the sum of complexities from all three modules, which is given by $\sum_{\ell=1}^{L_{\rm CorrMMF}+L_{\rm MMR}} \mathcal{O}(N C_{\ell-1} H_{\ell}W_{\ell}C_{\ell} K^{2}_{\ell})+\sum_{p=1}^{P_{\rm CSIR}}\mathcal{O}(N d_{p} h_{p})$, where the first term contains both the Corr-MMF module and MMR module due to their similar network structures.

\figref{time} further presents the inference time of 3D-CF construction for different methods. As observed, the proposed multimodal framework significantly reduces the inference time compared to the Kriging-based and GPR-based methods. 
It also achieves lower construction errors and enhanced flexibility with comparable time complexity compared with the GAN-based approach.
Relative to the state-of-the-art FL algorithm, the proposed scheme nearly halves the computational time.
It should be emphasized that the proposed 3D-CF is designed to be trained, inferred, and deployed at the BS side. Given the abundant computational resources available at the BS, this framework is practical.

To conclude, our multimodal framework exhibits competitive advantages in computational complexity, indicating great potential for practical applications.

\section{Conclusion} \label{conclusion}
In this paper, we proposed a modularized multimodal framework to construct 3D-CF for low-altitude communications.
Firstly, we established the 3D-CF model based on the ground-to-LAV Rician fading channels, which were defined as a collection of CSI-tuples with each tuple composed of LAV's positions and its corresponding channel information. Due to the heterogeneous structures of different prior data such as LAV coordinates, geographic environment maps, and sampling data, we transformed the 3D-CF construction problem into a multimodal regression task and proposed a high-efficiency modularized multimodal framework accordingly, where the Corr-MMF module and the MMR module were designed to extract features of CSI distribution in horizontal and vertical directions, and the CSI-R module was developed to estimate the target CSI and reconstruct the 3D-CF. Numerical results demonstrated the competitive performance and generalization ability of our proposed 3D-CF multimodal framework, which attains an accuracy improvement of at least 27.5\% over the benchmarks under different communication scenarios. We also analyzed the computational complexity and illustrated its superiority in terms of the inference time.
In the future, it will be interesting to quantify the impact of measurement data resolution on the constructed 3D-CF accuracy and its computational complexity. Furthermore, considering air-to-air channels and dynamic environmental factors also constitutes a highly promising research direction for low-altitude 3D-CF techniques.

\bibliographystyle{IEEEtran}
\bibliography{ref.bib}

\end{document}